%% file: ICTON_final_standalone/Main_Text.tex
\documentclass[10pt, a4paper, twoside, final, onecolumn]{icton}

\IEEEoverridecommandlockouts
\usepackage{amsmath,amssymb,dsfont,color,float,empheq}
\usepackage{graphicx,wrapfig,multicol,multirow}
\usepackage[dvipsnames]{xcolor}
\usepackage{psfrag,wrapfig}
\usepackage[font=footnotesize]{caption}
\usepackage{pgfplots}
\usepackage{subcaption}
\usepackage{cite}
\usepackage{setspace}
\usepackage{multicol}
\usetikzlibrary{arrows.meta}
\usetikzlibrary{patterns}
\usepackage{wrapfig}
\usepackage{makecell}
\usepgfplotslibrary{fillbetween}

\usepackage{tikz,pgfplots}
\usetikzlibrary{positioning,shadows,shapes,backgrounds,calc}
\usepackage{pgfplotstable}
\usepackage[free-standing-units]{siunitx}
\usepackage{circuitikz}
\pgfplotsset{compat=1.12} 

\definecolor{orange}{rgb}{1,0.7,0}
\definecolor{myDarkGreen}{rgb}{0.00000,0.58824,0.00000}%

\begin{document}

\title{Symbol Rate-Code Rate Trade-offs for IM/DD 200G/400G per Lane LPO Transceivers}

\author{{\authorblockN{{\bfseries José Núñez-Kasaneva$^1$, Yunus Can Gültekin$^1$, Stefanos Dris$^2$, Paraskevas Bakopoulos$^2$, Nikos Argyris$^2$, Gabriele Liga$^1$}}}\\
{\small \itshape {$^1$Department of Electrical Engineering,
Eindhoven University of Technology (TU/e), 5600MB Eindhoven, the Netherlands}}\\
\small \itshape {$^2$NVIDIA, Ermou 56, 10563 Athens, Greece}\\
\small \itshape {email:j.i.nunez.kasaneva@tue.nl}
}

% \author{José Núñez-Kasaneva\authormark{1,*}, Yunus Can Gültekin\authormark{1}, Stefanos Dris \authormark{2}, Paraskevas Bakopoulos \authormark{2}, Nikos Argyris\authormark{2}, and Gabriele Liga\authormark{1}}

% \address{\authormark{1} Department of Electrical Engineering, %Information and Communication Theory Laboratory, Signal Processing Systems Group,
% Eindhoven University of Technology, Eindhoven, the Netherlands
% \authormark{2}NVIDIA, Ermou 56, 10563 Athens, Greece}

% \email{\authormark{*}j.i.nunez.kasaneva@tue.nl} 

\maketitle
% \footnotetext[~]{979-8-3195-4420-9/26/\$31.00~\copyright~2026 IEEE}
% \IEEEpubid{979-8-3195-4420-9/26/\$31.00~\copyright~2026 IEEE}
\thispagestyle{empty}
\begingroup
\renewcommand{\thefootnote}{}
\footnotetext{ \centering 979-8-3195-4420-9/26/\$31.00~\copyright~2026 IEEE}
\endgroup

\begin{abstract}
We analyze the symbol rate--code rate trade-off in bandwidth-limited IM/DD systems targeting LPO transceivers, using PAM-4/6/8 as candidate modulation formats. We use the capacity of the binary symmetric channel as an achievable information rate under hard-decision decoding, serving as a performance metric for 200G and 400G per-lane throughput targets. 
Our results show that reducing the FEC code rate below the KP4 baseline allows higher-order PAM formats to operate at substantially lower symbol 
rates than PAM-4 while meeting the throughput requirement.
\end{abstract}
% \vspace{1mm}
% \begin{abstract}
% We study the information rate performance of bandlimited 200G and 400G per lane IM-DD systems. We highlight the trade-offs between FEC overhead, PAM modulation order, and the corresponding reduction in required symbol rates.
% \end{abstract}

\keywords{intensity-modulation/direct-detection (IM/DD), linear-drive pluggable optics (LPO), short-reach data-center interconnects (DCIs), Forward Error Correction (FEC).}

\vspace{-2mm}
\section{Introduction}
Artificial intelligence workloads are driving higher per-lane data rates in short-reach data-center interconnects (DCIs), where intensity-modulation/direct-detection (IM/DD) with pulse-amplitude modulation (PAM) formats remain attractive for their cost, complexity, and energy-efficiency advantages \cite{DI_CHE_IMDD_200}. Current commercial deployments have reached ~200 Gb/s per lane with PAM-4~\cite{st_arnaultNet32Tbps_OFC2025}. However, scaling to 400 Gb/s and beyond is mainly hindered by two constraints: (i) the bandwidth limitations in the transmission chain\cite{Pang_200G} and (ii) transmitter laser relative intensity noise (RIN)\cite{ agrawalFiberopticCommunicationSystems2002}. In particular, in IM/DD linear drive pluggable optics (LPO) transceivers, frequency-dependent loss arising from electronic components such as packages and printed circuit boards (PCBs), specifically the PCB traces that connect the optical module to the host digital signal processor, introduces strong intersymbol interference, especially at the high symbol rates required for 400 Gb/s/lane. PAM-4 remains the most widely adopted modulation format and is typically combined with precoding/equalization blocks as well as low-latency forward error correction (FEC) such as KP4, as defined in the IEEE 802.3dj standard~\cite{farhood_P802_3dj}. 
%%%%%%%%%%%%%%%%%%%%%%%%%%%%%%%%%%

%%%%%%%%%%%%%%%%%%%%%%%%%%%%%%%%%%
On the other hand, increasing the modulation order $M$ (e.g., PAM-6/8 \cite{PAM_6,PAM_8}) remains an attractive option, as it reduces the required symbol rate and, consequently, the DAC and ADC sampling rates, thereby lowering power consumption. However, higher-order PAM is generally believed to require a higher signal-to-noise ratio (SNR) at the receiver to achieve sufficiently low post-decoding error probabilities. 
This perception stems from the standard practice of fixing the FEC code rate (KP4) and using the pre-FEC bit error rate (BER) at the target baud rate as the pass/fail criterion against the KP4 threshold. This fixed-rate criterion penalizes higher-order PAM, which may not meet the KP4 threshold under severe bandwidth limitations and channel losses, even though higher modulation cardinalities can significantly reduce symbol rate and DAC/ADC power.
%%%%%%%%%%%%%%%%%%%%%%%%%%%%%%%%

%%%%%%%%%%%%%%%%%%%%%%%%%%%%%%%%
In this paper, we challenge this fixed-rate evaluation framework by characterizing IM/DD PAM-$4/6/8$ performance under variable code rate and symbol rate using achievable information rates (AIR) for hard-decision decoding. 
We show that pairing higher modulation orders with appropriately reduced code rates unlocks significant design flexibility, enabling 200G and 400G per-lane operation under stringent bandwidth constraints that would not be possible under a fixed-rate KP4 adoption.
% For each PAM-$M$ format, we determine the AIR and the maximum FEC code rate required to achieve error-free post-FEC operation under bandwidth limitation. Our results show that combining higher modulation orders with a lower code rates constraint offers additional design flexibility to achieve 200G and 400G per-lane performance under stringent bandwidth constraints. 
% \vfill
% \begin{center}
% \footnotesize{979-8-3195-4420-9/26/\$31.00 ©2026 IEEE}
% \end{center}

\vspace{-2mm}
\section{System Model}\label{Sec:SystemModel}

\input{Sections/2_Model}

\vspace{-2mm}

 \section{Results}
\label{sec:Results}
\input{Sections/3_Results}

\vspace{-2mm}

\section{Conclusions}
\label{sec:Conclusions}

We analyze the symbol rate--code rate trade-off for PAM-4/6/8 IM/DD systems targeting 200G and 400G per-lane LPO transceivers under stringent bandwidth constraints. Our results show that the standard fixed-rate KP4 
paradigm penalizes higher-order PAM formats, which cannot meet the KP4 threshold under severe bandwidth limitations despite offering significant symbol rate advantages over PAM-4. By instead characterizing performance via an achievable information rate under hard-decision decoding, we identify feasible symbol rate operating regions under variable-rate hard-decision FEC. 
Within these regions, we identify a code rate backoff margin that can accommodate practical FEC schemes for DCI applications. As a result, we show that a moderate FEC rate reduction compared to the KP4 
benchmark allows higher-order PAM formats to achieve substantial symbol rate reductions relative to PAM-4 at both 200G and 400G, unlocking significant design flexibility in the symbol rate and bandwidth allocation of 
future LPO transceivers. Future work will investigate the performance of pragmatic FEC schemes and their required code rate backoff, as well as the implementation complexity tradeoffs associated with lower FEC code rates and higher-order modulation formats.
\vspace{-2mm}

%adopting $R_\text{c} = 0.96$ and $R_\text{c} = 0.92$ for PAM-6 and PAM-8 enables symbol rate reductions of $20.8$ and $28$~GBd relative to PAM-4 at 200G, while at 400G, a backoff to $R_\text{c} = 0.91$ for PAM-8 yields a reduction of $54.1$~GBd.

%\setstretch{0.55}{\footnotesize

\section*{ACKNOWLEDGEMENTS}This project has received funding from the European Union’s Horizon Europe research and innovation program under the Marie Skłodowska-Curie grant agreement No 101119983.
\vspace{-1mm}

%\bc{Todos: Add reference and check}

% \bibliographystyle{IEEEtran}
% {\tiny
% \bibliography{IEEEabrv,sample}
% }

\begingroup
\scriptsize   % or \scriptsize (avoid \tiny)
\bibliographystyle{IEEEtran}
\bibliography{IEEEabrv,sample}
\endgroup

% }

\thispagestyle{empty}
\end{document}

%% file: Sections/2_Model.tex
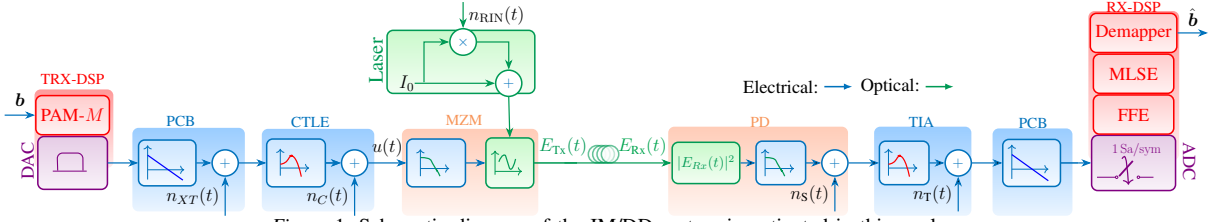
\begin{figure}[!t]
    \centering
    \resizebox{\textwidth}{!}{\input{Figures/schemes/New_E2E_LPO}}

    \caption{Schematic diagram of the IM/DD system investigated in this work.}
    \label{fig:E2E_combined}
    \vspace{-3mm}
\end{figure}
\begin{wrapfigure}[11]{R}{0.405\textwidth}
    % \clearpage
    \centering
    \resizebox{0.43\textwidth}{!}{%
        \input{Figures/plots/filters/all_filters}
    }
    \caption{Frequency responses of the individual TRX/RX filter blocks.}
    \label{fig:all_filters}
    % \vspace{-4mm}
\end{wrapfigure}
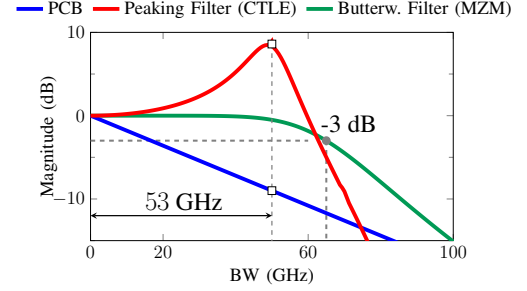

We model an IM/DD LPO system as depicted in Fig.~\ref{fig:E2E_combined}. At the 
transmitter, a bit sequence $\mathbf{b}$ is mapped to a PAM-$M$ symbol sequence ($M \in \{4,6,8\}$). The resulting symbols are upsampled by an ideal DAC  using rectangular pulse shaping, followed by a Gaussian filter to smooth the waveform before the electrical front-end. The transmitter frontend emulates the PCB 
channel via a linear loss filter, which imposes a fixed loss of $9$~dB at $53$~GHz 
\cite{luo800GLinearDirect} and whose frequency response is shown in 
Fig.~\ref{fig:all_filters}. PCB crosstalk is modeled as a colored zero-mean Gaussian 
process $n_{XT}(t)$ added at the PCB frequency response output.

% We model an IM/DD LPO system as depicted in Fig.~\ref{fig:E2E_combined}a. At the 
% transmitter, a bit sequence $\mathbf{b}$ is mapped to a PAM-$M$ symbol sequence ($M \in \{4,6,8\}$). The resulting symbols are upsampled by an ideal DAC  using rectangular pulse shaping, followed by a Gaussian filter to smooth the waveform before the electrical front-end. The transmitter frontend emulates the PCB 
% channel via a linear loss filter, which imposes a fixed loss of $9$~dB at $53$~GHz 
% \cite{luo800GLinearDirect} and whose frequency response is shown in 
% Fig.~\ref{fig:all_filters}. PCB crosstalk is modeled as a colored zero-mean Gaussian 
% process $n_{XT}(t)$ added at the PCB frequency response output.

The PCB loss is compensated by an active continuous-time linear equalizer (CTLE), modeled 
as a peaking filter using a second-order low-pass filter with a tunable damping factor, 
designed to fully restore the signal at the Nyquist frequency. The CTLE introduces 
additive noise $n_{C}(t)$, modeled as zero-mean white Gaussian noise. The noise powers of 
$n_{XT}(t)$ and $n_{C}(t)$ are set equal and jointly adjusted to yield a specified 
transmitter SNR (TRX-SNR) at the CTLE output $u(t)$.

The Mach–Zehnder modulator (MZM) is modeled using a 4th-order Butterworth filter, whose frequency response is 
shown in Fig.~\ref{fig:all_filters}, followed by the sinusoidal electro-optical transfer function. The MZM operates in push-pull configuration, 
driven by a continuous-wave laser with optical intensity $I_0$ and colored RIN $n_{\text{RIN}}(t)$ whose total power equals that of an 
equivalent white noise spectrum. The MZM bias voltage and drive amplitude determine the 
operating point, setting both the extinction ratio (ER)~\cite{DI_CHE_IMDD_200} and the 
optical modulation amplitude (OMA).

The resulting optical signal $E_{\text{TX}}(t)$ propagates over a single-mode fiber with 
$3$~dB attenuation before arriving at the receiver frontend. A photodiode (PD) performs 
square-law detection, followed by a 6th-order Butterworth filter modeling the photodiode 
frequency response. The detected current is amplified by a transimpedance amplifier 
(TIA), modeled as a peaking filter to pre-compensate the receiver PCB loss at the Nyquist 
frequency. Shot noise $n_{\text{S}}(t)$ and thermal noise $n_{\text{T}}(t)$ are added as 
zero-mean Gaussian processes, after which a PCB loss filter is applied.

Finally, the received signal is sampled at 1 Sa/sym via an ideal ADC assuming ideal time recovery and then processed by the receiver DSP chain. The intersymbol interference induced by bandwidth limitations in the system and additive noise is mitigated by a feed-forward equalizer (FFE) with 12 taps 
(8 pre-cursor, 3 post-cursor), followed by a 1-tap maximum-likelihood sequence estimator 
(MLSE). A hard symbol-to-bit demapper then recovers the estimated bit sequence $\mathbf{\hat{b}}$.

%% file: Figures/schemes/New_E2E_LPO.tex
\begin{tikzpicture}[
    line width = 1pt,
    every node/.style = {
        node distance = 2em,
        minimum width = 2em,
        minimum height = 2em
    },
    block/.style = {
        draw = RoyalBlue,
        text = RoyalBlue,
        shading = axis,
        top color = RoyalBlue!20!white,
        bottom color = RoyalBlue!5!white,
        rectangle, rounded corners, minimum width=1.1cm, minimum height=1cm, align=center, font=\large
    },
     highlightTRX/.style = {
         rounded corners,
         draw = none,
         text = Plum,
         shading = axis,
         top color =  RoyalBlue!40!white,
         bottom color =  RoyalBlue!8!white,
     },
      highlightTRX_2/.style = {
         rounded corners,
         draw = none,
         text = Plum,
         shading = axis,
         top color =  Peach!40!white,
         bottom color =  Peach!8!white,
     },
    redblock/.style = {
        draw = red,
        text = red,
        shading = axis,
        top color = red!20!white,
        bottom color = red!5!white,
        rectangle, rounded corners,
        minimum width=1.55cm,
        minimum height=0.9cm,
        align=center,
        font=\large
    },
    grayblock/.style = {
        draw = gray,
        text = gray,
        shading = axis,
        top color = gray!20!white,
        bottom color = gray!5!white,
        rectangle, rounded corners,
        minimum width=1.55cm,
        minimum height=0.9cm,
        align=center,
        font=\large
    },
    purpleblock/.style = {
        draw = violet,
        text = violet,
        shading = axis,
        top color = violet!20!white,
        bottom color = violet!5!white,
        rectangle, rounded corners,
        minimum width=1.1cm,
        minimum height=1cm,
        align=center,
        font=\large
    },
    orangeblock/.style = {
        draw = orange!80!black,
        text = orange!80!black,
        shading = axis,
        top color = orange!20!white,
        bottom color = orange!5!white,
        rectangle, rounded corners,
        minimum width=1.1cm,
        minimum height=1cm,
        align=center,
        font=\large
    },
    greenblock/.style = {
        draw = ForestGreen,
        text = ForestGreen,
        shading = axis,
        top color = ForestGreen!20!white,
        bottom color = ForestGreen!5!white,
        rectangle, rounded corners,
        minimum width=1.1cm,
        minimum height=1cm,
        align=center,
        font=\large
    },
    ellipseblock/.style={circle, draw = RoyalBlue, fill = white, text = RoyalBlue, minimum size=0.5cm, font=\footnotesize},
    addition/.style={circle, draw = RoyalBlue, fill = white,scale=0.8, text = RoyalBlue, minimum size=0.5cm, font=\footnotesize},
    amp/.style={draw = RoyalBlue, isosceles triangle, isosceles triangle apex angle=60, fill=white, 
    minimum height=1cm, minimum width=1cm, rotate=0, anchor=center},
    amp1/.style={draw = RoyalBlue, isosceles triangle, isosceles triangle apex angle=60, fill=white, 
    minimum height=1cm, minimum width=1cm, rotate=180, anchor=center},
    arrow/.style={-{Stealth[scale=1]}, thick, draw = RoyalBlue},
    backgroundBlockRed/.style={draw = red,
    text = red,
    shading = axis,
    top color = red!20!white,
    bottom color = red!5!white,, rounded corners, minimum width=6cm, minimum height=2.5cm},
    backgroundBlock/.style={draw = RoyalBlue!40!black, fill = RoyalBlue!5, rounded corners, minimum width=6cm, minimum height=2.5cm},
    highlight/.style = {
        rounded corners,
        draw = none,
        text = Plum,
        shading = axis,
        top color = red!40!white,
        bottom color = red!8!white
    }
]

% \node[purpleblock, scale=0.9, align=center,minimum width=1.78cm,rotate=0] (preeq) {$\uparrow$};

\node[purpleblock, scale=0.9, align=center,
      minimum width=1.78cm, minimum height=1.35cm,
      rotate=0] (preeq) {    
\begin{tikzpicture}[scale=0.28, baseline=-0.3ex]
    % small axes
    \draw[violet, thin] (-0.2,0) -- (2.9,0);
    % \draw[violet, thin] (0,0) -- (0,1.2);

    % rectangular pulse shape
    \draw[violet, thick]
        (0.25,0.05) -- (0.2,1.65)
        -- (2.55,1.65) -- (2.56,0.05);

    % rotated label inside
    % \node[rotate=90, anchor=center] at (-0.5,0.8) {DAC};
\end{tikzpicture} 
};

 \node[text = violet, above =0.1em of preeq.north , xshift=-1.9em, yshift=-1.7em, align = center,rotate=90,scale=1] (DACC) {\large DAC};

% \node[ redblock, above=0.05em of preeq, align=center] (pshape) {Pulse\\ Shaping};
\node[ redblock, above=0.05em of preeq, align=center] (Mod) {PAM-$M$};

 \node[text = red, above =2.7em of preeq.north , align = center] (Diversity1) {{TRX-DSP}};
\begin{scope}[on background layer]
\draw[highlight] ($(Diversity1.north west) + (-0.01,-0.61)$) rectangle ($(Diversity1.east |- preeq.south) + (0.07, .5)$);
\end{scope}

\draw[->, arrow] ($(Mod.west)+(-0.7,0)$) -- ($(Mod.west)+(0,0)$) node[midway,above] {\large $\boldsymbol{b}$};

% \draw[->, arrow] (PF.east) -- (AMP.west);
%%%%%%%%%%%%%%%%%%%%%%%%%%%%%%%%%%%%%%%%%%%%%%%%%%%%%%%%%%%%%%%TX_FON%%%%%%%%%%%%%%%%%%%%%%%%%%%%%%%%%%%%%%%%%%%%%%%%%%%%%%%%%%%%
    \node[block, right=2em of preeq,draw=RoyalBlue,
      minimum width=1.25cm, minimum height=1cm] (PCB) {
    \begin{tikzpicture}[scale=0.28]
        \draw[->, RoyalBlue] (0,0) -- (3,0);
        \draw[->, RoyalBlue] (0,-0.8) -- (0,2);
        \draw[blue, thick] (0,1.2) -- (2.9,-.85);
    \end{tikzpicture}
    % \\[-0.2em] 
    % TRF
    };

    % Amplifier addition
    \node[addition, right=1em of PCB] (add1) {\large $+$};

%% PCB Highlight
\node[above=2em of PCB.south east] (PCBbox) {};

\node[text=RoyalBlue, align=center, anchor=south] 
  at ($(PCBbox.north)+(-1em,-0.9em)$) {PCB};
    \begin{scope}[on background layer]
    \draw[highlightTRX] ($(PCBbox.north west) + (-1.1,-.1)$) rectangle ($(PCBbox.east |- add1.south) + (0.75, -0.8)$);
    \end{scope}

\node[block, right=1.8em of add1, draw=RoyalBlue,
      minimum width=1.35cm, minimum height=1.1cm] (PF) {
\begin{tikzpicture}[scale=0.22, baseline=-0.3ex]

    % Axes
    \draw[->, RoyalBlue, thin] (0,0) -- (3,0);
   \draw[->, RoyalBlue] (0,-0.8) -- (0,2);

    % Smooth CTLE response (Bézier)
    \draw[red, thick]
        (0,0.4)
        .. controls (0.8,0.5) and (0.85,1.1)
        .. (1.6,1.456)
        .. controls (1.4,1.4) and (1.8,0.9)
        .. (2.2,-0.9);
\end{tikzpicture}
% \\[-0.4em]
% {\small CTLE}
};

    % Amplifier addition
    \node[addition, right=1em of PF] (AMP) {\large $+$};

%% CTLE Highlight
\node[above=2em of PF.south east] (PF_TRX) {};

\node[text=RoyalBlue, align=center, anchor=south] 
  at ($(PF_TRX.north)+(-1em,-0.9em)$) {CTLE};
    \begin{scope}[on background layer]
    \draw[highlightTRX] ($(PF_TRX.north west) + (-1.1,-.1)$) rectangle ($(PF_TRX.east |- AMP.south) + (0.75, -0.8)$);
    \end{scope}

\node[block, right=2.5em of AMP,   minimum width=1.35cm, minimum height=1.1cm] (BTW){
\begin{tikzpicture}[scale=0.22, baseline=-0.3ex]

    % Axes
    \draw[->, RoyalBlue, thin] (0,0) -- (3,0);
   \draw[->, RoyalBlue] (0,-0.8) -- (0,2);

    % 4th-order Butterworth-like response
    \draw[ForestGreen, thick]
        (0,1.15)
        .. controls (0.6,1.13) and (1.0,1.1)
        .. (1.42,0.21)
        .. controls (1.5,0.1) and (1.85,-.3)
        .. (2.05,-0.8);

\end{tikzpicture}
};

%%MZM Highlight
\node[above=2em of BTW.south east] (BTW_TRX) {};

\node[text=Peach, align=center, anchor=south] 
  at ($(BTW_TRX.north)+(-0.1em,-.75em)$) {MZM};
    \begin{scope}[on background layer]
    \draw[highlightTRX_2] ($(BTW_TRX.north west) + (-1.1,-.1)$) rectangle ($(BTW_TRX.east |- AMP.south) + (1.3, -.9)$);
    \end{scope}

     \node[greenblock, right=0.3em of BTW,   minimum height=1.15cm,draw=ForestGreen] (IM) at ($(BTW.east)+(1em,0)$) {

\begin{tikzpicture}[scale=0.22, baseline=-0.4ex]

    % Axes
    \draw[->, ForestGreen, thin] (-.5,-.85) -- (2.3,-0.85);
    \draw[->, ForestGreen, thin] (-.5,-1.2) -- (-.5,1.3);

    % Cosine transfer
     \draw[ForestGreen, thick]
        (-0.10,0.15)
        .. controls (0.20,0.95) and (0.55,0.95)
        .. (0.80,0.15)
        .. controls (1.05,-0.65) and (1.30,-0.95)
        .. (1.65,0.45);
        % .. controls (1.90,0.85) and (2.15,0.65);
        % .. (2.35,0.35);

        % \draw[ForestGreen, thick] (-0.1,0.6) .. controls (0.75,1.35) and (0.95,-2.5) .. (1.55,1.055) .. controls (1.75,0.85) and (2.1,0.45) .. (2.15,0);

\end{tikzpicture}
};

 % Laser/RIN block
\node[greenblock, above=1cm of IM, xshift=-3.2em,
      minimum width=3.4cm, minimum height=1.45cm,
      draw=ForestGreen] (RIN) {};

% Label
% \node[text=ForestGreen, font=\large] 
    % at ($(RIN.north)+(0,-0.18)$) {Laser};

% Internal nodes
\node[addition, draw=ForestGreen, scale=1] 
    (Nrin) at ($(RIN.south)+(1.1,0.25)$) {\large $+$};

\node[addition, draw=ForestGreen, scale=1] 
    (MultRIN) at ($(Nrin.center)+(-1.05,0.95)$) {\large $\times$};

% I0 vertical branch
\draw[->, arrow,ForestGreen, thick]
    ($(Nrin.west)+(-1.85,0)$)
    -- ($(Nrin.west)+(0,0)$)
    node[pos=0.1,left,black] {\large $I_0$};

% branch to multiplier
\draw[->,,arrow,ForestGreen, thick]
    ($(Nrin.north)+(-1.95,-0.29)$) |- (MultRIN.west);

% RIN input
\draw[->,arrow, ForestGreen, thick]
    ($(MultRIN.north)+(0,0.6)$)
    -- (MultRIN.north)
    node[midway,right,black] {\large $n_{\mathrm{RIN}}(t)$};

% multiplier output to adder
\draw[->, arrow,ForestGreen, thick]
    (MultRIN.east) -| (Nrin.north);

% output to MZM

     \node[text = ForestGreen, above =0.7em of Nrin.north , xshift=-7.5em, yshift=-0.1em, align = center,rotate=90,scale=1.12
     ] (DRIN) {\large Laser};

    % Connections
    \draw[->, arrow] (PCB.east) -- (add1.west);

    \draw[->, arrow] (add1.east) -- (PF.west);
    % \draw[->, arrow] (addXT.east) -- (PF.west);
    \draw[->, arrow] (PF.east) -- (AMP.west);

    \draw[->, arrow] (AMP.east) -- (BTW.west)node[midway,above] {\large $u(t)$};

    \draw[->, arrow] (PF.east) -- (AMP.west);
    % \draw[->, arrow] (add.east) -- (IM.west);
    %\draw[->, arrow] (add.east) -- (BTF.west);
    \draw[->, arrow] (BTW.east) -- (IM.west);
    \draw[->, arrow,ForestGreen] (Nrin.south) -- (IM.north);

    \draw[->, arrow] ($(add1.south)+(0,-0.9)$) -- ($(add1.south)$) node[midway,left] {\large $n_{XT}(t)$};

    \draw[->, arrow] ($(AMP.south)+(0,-0.9)$) -- ($(AMP.south)$) node[midway,left] {\large $n_{C}(t)$};
   
    \draw[->, arrow] (preeq.east) -- (PCB.west)node[midway,above]{};% {$\boldsymbol{s}_{\text{u}}(t)$};

    % \draw[->, arrow] (IM.east) -- (output.west);

  % Reference node (stays put and keeps your rectangle math unchanged)
\node[above=3em of PF.south east] (Diversity) {};

%% TRX-Frontend
% \node[text=RoyalBlue, align=center, anchor=south] 
%   at ($(Diversity.north)+(-7em,0)$) {TRX-Frontend};
%     \begin{scope}[on background layer]
%     \draw[highlightTRX] ($(Diversity.north west) + (-3.8,1.1)$) rectangle ($(Diversity.east |- AMP.south) + (2.55, -0.8)$);
%     \end{scope}

    %%%%%%%%%%%%%%%%%%%%%%% channel %%%%%%%%%%%%%%%%%%%%%%%%%%%%%%%%%%%%%%%%%%%%%%%%%%%%%%%

     %%%%%%%%%%%%%%%%%%%%%%% receiver frontend  %%%%%%%%%%%%%%%%%%%%%%%%%%%%%%%%%%%%%%%%%%%%%%%%%%%%%%%

    \node[greenblock, right=9em of IM,scale=0.95,draw=ForestGreen] (normRX) {\normalsize $\lvert E_{Rx}(t)\rvert^2$};

    % butterwort filter
    \node[block, right=1em of normRX,draw=RoyalBlue] (PCB_RX) {
\begin{tikzpicture}[scale=0.22, baseline=-0.3ex]

    % Axes
    \draw[->, RoyalBlue, thin] (0,0) -- (3,0);
   \draw[->, RoyalBlue] (0,-0.8) -- (0,2);

     % 4th-order Butterworth-like response
    \draw[ForestGreen, thick]
        (0,1.15)
        .. controls (0.6,1.13) and (1.0,1.1)
        .. (1.42,0.21)
        .. controls (1.5,0.1) and (1.85,-.3)
        .. (2.05,-0.8);

\end{tikzpicture}
};

    % shot noise addition
    \node[addition, right=1em of PCB_RX] (add_RX) {\large $+$};
    % Normalization

%%PD Highlight
\node[above=2em of PCB_RX.south east] (PD_RX) {};

\node[text=Peach, align=center, anchor=south] 
  at ($(PD_RX.north)+(-3em,-1em)$) {PD};
    \begin{scope}[on background layer]
    \draw[highlightTRX_2] ($(PD_RX.north west) + (-2.8,-.1)$) rectangle ($(PD_RX.east |- AMP.south) + (0.6, -.9)$);
    \end{scope}

    % Peaking Filter
    \node[block, right=2em of add_RX,draw=RoyalBlue] (PF_RX) {
\begin{tikzpicture}[scale=0.22, baseline=-0.3ex]

    % Axes
    \draw[->, RoyalBlue, thin] (0,0) -- (3,0);
   \draw[->, RoyalBlue] (0,-0.8) -- (0,2);

    % Smooth CTLE response (Bézier)
    \draw[red, thick]
        (0,0.4)
        .. controls (0.8,0.5) and (0.85,1.1)
        .. (1.6,1.456)
        .. controls (1.4,1.4) and (1.8,0.9)
        .. (2.2,-0.9);
\end{tikzpicture}
};

    % thermal noise addition
    \node[addition, right=1em of PF_RX] (TN) {\large $+$};

%%CTLE RX Highlight
\node[above=2em of PF_RX.south east] (PKF_RX) {};

\node[text=RoyalBlue, align=center, anchor=south] 
  at ($(PKF_RX.north)+(-0.5em,-1em)$) {TIA};
    \begin{scope}[on background layer]
    \draw[highlightTRX] ($(PKF_RX.north west) + (-.9,-.1)$) rectangle ($(PKF_RX.east |- AMP.south) + (0.6, -.9)$);
    \end{scope}

    % Linear Filtering
    \node[block, right=2em of TN,draw=RoyalBlue] (LF_RX) { \begin{tikzpicture}[scale=0.28]
        \draw[->, RoyalBlue] (0,0) -- (3,0);
        \draw[->, RoyalBlue] (0,-0.8) -- (0,2);
        \draw[blue, thick] (0,1.2) -- (3,-0.95);
    \end{tikzpicture}
    };

    %%CTLE RX Highlight
\node[above=2em of LF_RX.south east] (TRF_RX) {};

\node[text=RoyalBlue, align=center, anchor=south] 
  at ($(TRF_RX.north)+(-1.8em,-1em)$) {PCB};
    \begin{scope}[on background layer]
    \draw[highlightTRX] ($(TRF_RX.north west) + (-1.2,-.1)$) rectangle ($(TRF_RX.east |- AMP.south) + (-0.1, -.9)$);
    \end{scope}

    \draw[->, arrow] (normRX.east) -- (PCB_RX.west);
    \draw[->, arrow] (PCB_RX.east) -- (add_RX.west);
    \draw[->, arrow] (add_RX.east) -- (PF_RX.west);
    \draw[->, arrow] (PF_RX.east) -- (TN.west);
    \draw[->, arrow] (TN.east) -- (LF_RX.west);

     \draw[->, arrow] ($(TN.south)+(0,-0.8)$) -- ($(TN.south)$) node[midway,left] {\large $n_{\text{T}}(t)$};
    \draw[->, arrow] ($(add_RX.south)+(0,-0.8)$) -- ($(add_RX.south)+(0,0)$) node[midway,left] {\large $n_{\text{S}}(t)$};   

    % Draw the first horizontal segment and label
    % Original horizontal line + label
    \draw[-, ForestGreen]   ($(IM.east)+(0.0,0)$) 
                       -- ($(IM.east)+(0.99,0)$) 
                       node[midway,xshift=.5em, above, ForestGreen] {\large $E_{\text{Tx}}(t)$};
    
    % Ellipses over the fiber segment
    % Rising ellipses over the arrow path to simulate fiber
    % Ellipses over the fiber segment
    \foreach \x in {1.45,1.55,1.65,1.75} {
        \node[ellipse, draw=ForestGreen, fill=ForestGreen!1, minimum width=0.4cm, minimum height=0.4cm,opacity=0.6] 
            at ($(IM.east)+(\x, 0.2)$) {};
    }
    
    % Arrow continues from the last ellipse to normRX
    \draw[->, arrow,ForestGreen]   ($(IM.east)+(0.7,0)$) 
                   -- ($(normRX.west)+(0,0)$) node[midway, above, below=-20pt, xshift=1.6em, yshift=1, Green] {\large $E_{\text{Rx}}(t)$};;

    %  \node[text = RoyalBlue, below =1.3em of PF_RX.south, align = center] (Diversity2) {RX-Frontend};
    % \begin{scope}[on background layer]
    % \draw[highlightTRX] ($(Diversity2.north west) + (-4.1,2.5)$) rectangle ($(Diversity2.east |- PF_RX.south) + (2.3, -1.2)$);
    % \end{scope}

     \node[above=2.5em of PCB_RX] (ele) {\large Electrical:};

    % simple arrow to the right of the label
    \draw[-{Latex[length=2mm]},RoyalBlue] (ele.east) -- ++(0.65,0);
    
    \node[right=2em of ele] (opt) {\large Optical:};
    \draw[-{Latex[length=2mm]},ForestGreen] (opt.east) -- ++(0.65,0);
    %%%%%%%%%%%%%%%%%%%%%%% receiver DSP %%%%%%%%%%%%%%%%%%%%%%%%%%%%%%%%%%%%%%%%%%%%%%%%%%%%%%%
    %%%%%%%%%%%%%%%%%%%%%%% receiver DSP %%%%%%%%%%%%%%%%%%%%%%%%%%%%%%%%%%%%%%%%%%%%%%%%%%%%%%%

% Inverse precoding block
% \node[purpleblock, right=2em of LF_RX, yshift=-3.4em,minimum width=1.85cm] (Invprecod) {ADC};
\node[purpleblock, right=2em of LF_RX, yshift=0em,
      minimum width=1.35cm, minimum height=0.85cm] (Invprecod) {
\begin{circuitikz}[scale=0.5, transform shape]
\node[font=\Large] at (0.55,1.25) {$1\,\mathrm{Sa/sym}$};
\draw (0,0) to[short,-o] (0.85,0);
% \draw (0.85,0) to[short] (1.25,0);
\draw (1.85,0) to[short,o-] (2.60,0);
\draw[very thick] (1.02,0.05) -- (1.65,0.75);
\draw[thick,arrow,violet,->] (1.15,0.85)
  .. controls (1.3,1.05) and (1.35,1.10) .. (1.45,-0.35);
\end{circuitikz}
};

 \node[text = violet, above =0.1em of Invprecod.north , xshift=2.6em, yshift=-1.8em, align = center,rotate=-90,scale=1] (ADCC) {\large ADC};

\node[redblock, above=0.01em of Invprecod, yshift=0em,minimum width=1.85cm] (Eq) {FFE};
\node[redblock, above=0.01em of Eq, yshift=0em,minimum width=1.85cm] (MLSE) {MLSE};
\node[redblock, above=0.01em of MLSE,minimum width=1.85cm] (Demod) {Demapper};

% % Demodulator block
% \node[redblock, left=1.5em of Invprecod] (Demod) {Demodulator};

% Connections
\draw[-,RoyalBlue] ($(LF_RX.east)+(0,0)$)-- ($(LF_RX.east)+(0.4,0)$);
\draw[->, arrow] ($(LF_RX.east)+(0.4,0)$)|- ($(Invprecod.west)+(0,0)$);

    \draw[->, arrow] ($(Demod.east)+(0,0)$) -- ($(Demod.east)+(0.7,0)$) node[midway,above] {\large $\hat{\boldsymbol{b}}$};

    \node[text = red, above =11em of Invprecod.south, align = center] (Diversity3) {RX-DSP};
    \begin{scope}[on background layer]
    \draw[highlight] ($(Diversity3.north west) + (-0.28,-3.9)$) rectangle ($(Diversity3.east |- PF_RX.south) + (0.29, 4)$);
    \end{scope}

% \begin{scope}[on background layer]
%     \filldraw[gray, dashed, thick, rounded corners, fill=gray, fill opacity=0.2] 
%         ($(DFE.north west) + (-0.15,0.25)$) rectangle 
%         ($(MLSE.south east) + (0.25,-0.15)$);
% \end{scope}
% \node [font=\Large,anchor=north west] at (1.5,2) {\shortstack[l]{\textbf{(a)}}};
    
\end{tikzpicture}

%% file: Figures/plots/filters/all_filters.tex
\begin{tikzpicture}
\begin{axis}[
    width=9.8cm, 
    height=6.3cm,
    xlabel={\large BW (GHz)},
    ylabel={\large Magnitude (dB)},
    ylabel shift = -10pt,
    xmin=0, xmax=100,
    ymax=10,
    ymin=-15,
    xtick={0, 20, 60, 100},
    minor y tick num=1,
    tick label style={font=\large},
    legend style={
        at={(0.48,1.18)}, % below the plot
        anchor=north,
        legend columns=3,
        /tikz/every even column/.append style={column sep=3pt},
        font=\large,
        draw=none,
        fill=none
    },
]
\addlegendimage{blue, line width=2.5pt}
\addlegendentry{PCB}
\addlegendimage{red, line width=2.5pt}
\addlegendentry{Peaking Filter (CTLE)}
\addlegendimage{ForestGreen, line width=2.5pt}
\addlegendentry{Butterw. Filter (MZM)}

% Draw -3dB horizontal line
\draw[dashed, line width=1.3pt,gray] (axis cs:0,-3) -- (axis cs:60,-3);
% Draw cutoff frequency vertical line
\draw[dashed, line width=1.3pt,gray] (axis cs:65,-3) -- (axis cs:65,-80);

% 4th-order Butterworth filter response
\addplot[domain=0:140, samples=100, smooth, ForestGreen, line width=2.5pt] {
    -10*log10(1 + (x/65)^8)  % 2n=8 for 4th-order
};
% \addlegendentry{$n=4$}

% Add cutoff point label
\node[circle, fill=gray, inner sep=2pt] at (axis cs:65,-3) {};
\node[anchor=west] at (axis cs:62,-1) {\Large -3 dB};

\addplot[mark=square*,  black ,mark options={solid, fill=white}, only marks, mark size=2.5pt] coordinates {(50, 8.6)}; 
% , mark options={solid, fill=white}
% Optional: Mark Rs/2 and Fs/2
% \draw[dashed, thick, gray] (axis cs:53,-40) -- (axis cs:53,50);

% \draw[<->,style={draw,>=stealth,rounded corners},thick,black] (axis cs:0,-15.0025) -- (axis cs: 53,-15) node[midway,above=0pt, yshift=-.02cm, xshift=-.09cm] { $R_s/2$};

%%%%%%%%%%%%%%%%%%%%%%%%%%%%%%%%%PCB%%%%%%%%%%%%%%%%%%%%%%%%%%%%%%%%%%%%%%%%%%%%%%%%%%%%%%%%%%%%%%%%%%
%%%%%%%%%%%%%%%%%%%%%%%%%%%%%%%%%PCB%%%%%%%%%%%%%%%%%%%%%%%%%%%%%%%%%%%%%%%%%%%%%%%%%%%%%%%%%%%%%%%%%%
%%%%%%%%%%%%%%%%%%%%%%%%%%%%%%%%%PCB%%%%%%%%%%%%%%%%%%%%%%%%%%%%%%%%%%%%%%%%%%%%%%%%%%%%%%%%%%%%%%%%%%
%%%%%%%%%%%%%%%%%%%%%%%%%%%%%%%%%PCB%%%%%%%%%%%%%%%%%%%%%%%%%%%%%%%%%%%%%%%%%%%%%%%%%%%%%%%%%%%%%%%%%%

\addplot[blue, line width=2.5pt] table[x=f, y=H, col sep=space]{Figures/plots/filters/plot_linear_filter_9dB.txt};\label{L9dB}
%\addlegendentry{Linear Filter $-9$~dB}

% \addplot[ForestGreen, dashed, very thick] table[x=f, y=H, col sep=space]{data/WP3/plot_linear_filter_20.txt};\label{L20dB}
%\addlegendentry{Linear Filter $-20$~dB}
% Dashed line representi
\draw[dashed, thick, gray] (axis cs:50,-40) -- (axis cs:50,9);

%Adding pooint
\addplot[mark=square*,  black ,mark options={solid, fill=white}, only marks, mark size=2.5pt] coordinates {(50, -9)}; 
% \addplot[mark=square*black ,mark options={solid, fill=white}, only marks] coordinates {(50, -9)}; 

\draw[<->,style={draw,>=stealth,rounded corners},thick,black] (axis cs:0,-12.0025) -- (axis cs: 50,-12) node[midway,above=0pt, yshift=-.02cm, xshift=.06cm] { \Large $53$~GHz};

%%%%%%%%%%%%%%%%%%%%%%%%%%%%%%%%%Peaking%%%%%%%%%%%%%%%%%%%%%%%%%%%%%%%%%%%%%%%%%%%%%%%%%%%%%%%%%%%%%%%%%%
%%%%%%%%%%%%%%%%%%%%%%%%%%%%%%%%%Peaking%%%%%%%%%%%%%%%%%%%%%%%%%%%%%%%%%%%%%%%%%%%%%%%%%%%%%%%%%%%%%%%%%%
%%%%%%%%%%%%%%%%%%%%%%%%%%%%%%%%%peaking%%%%%%%%%%%%%%%%%%%%%%%%%%%%%%%%%%%%%%%%%%%%%%%%%%%%%%%%%%%%%%%%%%
%%%%%%%%%%%%%%%%%%%%%%%%%%%%%%%%%peaking%%%%%%%%%%%%%%%%%%%%%%%%%%%%%%%%%%%%%%%%%%%%%%%%%%%%%%%%%%%%%%%%%%

\addplot [red, line width=2.5pt]
  table[row sep=crcr]{%
0	1.92865493310657e-15\\
0.625	0.00077573419441332\\
1.25	0.00310388237047175\\
1.875	0.00698728400303075\\
2.5	0.0124306805516578\\
3.125	0.0194407290099869\\
3.75	0.0280260209693844\\
4.375	0.0381971072922966\\
5	0.0499665285189596\\
5.625	0.0633488511603406\\
6.25	0.0783607100603362\\
6.875	0.0950208570417063\\
7.5	0.113350216082998\\
8.125	0.133371945308177\\
8.75	0.155111506106894\\
9.375	0.178596739741696\\
10	0.203857951838842\\
10.625	0.230928005202523\\
11.25	0.259842421437676\\
11.875	0.290639491915054\\
12.5	0.323360398663457\\
13.125	0.358049345828429\\
13.75	0.394753702394126\\
14.375	0.433524156925608\\
15	0.474414885152295\\
15.625	0.517483731279365\\
16.25	0.562792403982193\\
16.875	0.610406688108516\\
17.5	0.660396673182912\\
18.125	0.71283699987672\\
18.75	0.767807125671354\\
19.375	0.825391611001224\\
20	0.885680427209491\\
20.625	0.948769287680481\\
21.25	1.01476000351864\\
21.875	1.08376086511544\\
22.5	1.15588705086892\\
23.125	1.23126106417756\\
23.75	1.31001319959685\\
24.375	1.39228203869141\\
25	1.47821497559581\\
25.625	1.56796877155755\\
26.25	1.66171013670112\\
26.875	1.75961633582593\\
27.5	1.86187581310011\\
28.125	1.96868882786495\\
28.75	2.08026809018682\\
29.375	2.19683937997586\\
30	2.31864212701785\\
30.625	2.44592992058064\\
31.25	2.57897090561259\\
31.875	2.71804800694171\\
32.5	2.86345890195828\\
33.125	3.01551563419784\\
33.75	3.17454372258201\\
34.375	3.34088057051929\\
35	3.514872911191\\
35.625	3.69687293423447\\
36.25	3.88723261685842\\
36.875	4.08629561893618\\
37.5	4.29438588367174\\
38.125	4.51179179665384\\
38.75	4.73874437716217\\
39.375	4.97538748569134\\
40	5.22173741293274\\
40.625	5.47762846270316\\
41.25	5.74264028201743\\
41.875	6.01600182383309\\
42.5	6.29646618839616\\
43.125	6.58215066381447\\
43.75	6.87033797625155\\
44.375	7.15723957905225\\
45	7.43773201349864\\
45.625	7.70509571878455\\
46.25	7.95081419086271\\
46.875	8.16452830290489\\
47.5	8.33427489162759\\
48.125	8.44714466957741\\
48.75	8.49043416512518\\
49.375	8.54321223112171\\
50	8.9280010125883\\
50.625	8.11209622346709\\
51.25	7.80807058392426\\
51.875	7.42327012915719\\
52.5	6.96849367755777\\
53.125	6.45631160553185\\
53.75	5.89949907772113\\
54.375	5.3098838409942\\
55	4.69768803994962\\
55.625	4.07128935082251\\
56.25	3.43726637227242\\
56.875	2.8005991672896\\
57.5	2.16493014187163\\
58.125	1.5328273590947\\
58.75	0.906020909137604\\
59.375	0.285601300553134\\
60	-0.327820958705803\\
60.625	-0.933990603281415\\
61.25	-1.53291199866704\\
61.875	-2.12477692452908\\
62.5	-2.70991064179968\\
63.125	-3.288732000258\\
63.75	-3.86172420300273\\
64.375	-4.42941359741963\\
65	-4.99235448149007\\
65.625	-5.55111840477624\\
66.25	-6.10628682305195\\
66.875	-6.65844625511167\\
67.5	-7.2081853091472\\
68.125	-7.75609311061848\\
68.75	-8.30275878693793\\
69.375	-8.64877175681376\\
70	-9.09472264162239\\
% 70.625	-9.3412046686618\\
71.25	-10.4888154760559\\
71.875	-11.0381592598033\\
72.5	-11.5898492275257\\
73.125	-12.1445103427937\\
73.75	-12.7027823599667\\
74.375	-13.2653231634329\\
75	-13.8328124379187\\
75.625	-14.4059557089382\\
76.25	-14.9854888051966\\
76.875	-15.5721828085305\\
77.5	-16.1668495724768\\
78.125	-16.7703479086084\\
78.75	-17.3835905612824\\
79.375	-18.0075521175547\\
80	-18.6432780311494\\
80.625	-19.291894979381\\
81.25	-19.9546228222167\\
81.875	-20.6327884964592\\
82.5	-21.3278422596332\\
83.125	-22.0413768034494\\
83.75	-22.775149893792\\
84.375	-23.5311113742608\\
85	-24.311435609254\\
85.625	-25.1185607629742\\
86.25	-25.9552367452186\\
86.875	-26.8245842511965\\
87.5	-27.730168152123\\
88.125	-28.6760896637326\\
88.75	-29.6671033973348\\
89.375	-30.7087678439391\\
90	-31.8076414764844\\
90.625	-32.9715421703616\\
91.25	-34.2098962091359\\
91.875	-35.5342168021774\\
92.5	-36.9587744756253\\
93.125	-38.5015598072158\\
93.75	-40.1857062679951\\
94.375	-42.0416652383123\\
95	-44.110667426686\\
95.625	-46.4505077597172\\
96.25	-49.1458198056456\\
96.875	-52.3278026250766\\
97.5	-56.2162408544002\\
98.125	-61.223142710238\\
98.75	-68.2734678596622\\
99.375	-80.3186703643791\\
};\label{zeta1779}

% \node [font=\huge,anchor=north west] at (70,10) {\shortstack[l]{\textbf{(b)}}};

\end{axis}
\end{tikzpicture}

%% file: Sections/3_Results.tex
\subsection{Achievable Information Rate}

In this work, we adopt AIRs to analyze the performance of a coded IM/DD transmission system. An AIR is a transmission rate (but not necessarily the maximum) at which reliable transmission can be achieved for a given channel and transmitter-receiver processing pair, via a sufficiently strong channel code. 
%An AIR indicates the information rate at which error-free transmission is possible for an optimal code and decoder~\cite{AIR_alex}. 
Specifically, here, we compute AIRs for uniform PAM-$M$ constellations with (ideal) hard-decision (HD) binary decoders. Let $M$ be the PAM order, $m=\log_2 M$ be the number of bits mapped to each PAM symbol, and $p$ is the average HD BER measured at the output of the hard demapper in the RX-DSP in Fig.~\ref{fig:E2E_combined}. Then, an AIR in bits per symbol is given by the quantity \cite[Sec.~III-C]{achievable_rate_HD}
% \vspace{-2mm}
\begin{equation}
    {AIR}_{\mathrm{HD}}=m\left[1-\mathbb{ H}_2(p)\right],  \label{eq: AIR}\nonumber
\end{equation}
% \begin{align}
%    {AIR}_{\mathrm{HD}}=m\left[1-\mathbb{ H}_2(p)\right],  \label{eq: AIR}\nonumber
% \end{align}
%\end{minipage}
%\hfill
%\begin{minipage}{0.45\textwidth}  % Segunda columna ocupando el 45% del ancho
%\end{minipage}
where 
$\mathbb{H}_2(p)=-p\log_2p-(1-p)\log_2(1-p)$ is the binary entropy function. The corresponding achievable throughput (in bit/s) is $T_{\mathrm{AIR}}=R_s\cdot {AIR}_{\mathrm{HD}}$ where $R_s$ denotes the symbol rate.

\subsection{Numerical Results}
 
In this section, we numerically compare the performance of PAM-$M$ formats for $200$ and $400$~Gb/s transmission, sweeping $B_{\text{MZM}}$ across the 45-65~GHz and 100-120~GHz ranges, respectively, using the model described in Sec.~\ref{Sec:SystemModel} and shown in Fig.~\ref{fig:E2E_combined}. The simulation parameters are listed in Table~\ref{tab:fixed_code_rate}.

% \begin{table}[t]
% \centering
% \small
% \caption{Simulation Parameters Used for the Numerical Results}
% \label{tab:fixed_code_rate}
 
% \setlength{\tabcolsep}{1pt}
% \renewcommand{\arraystretch}{0.80}
 
% \begin{tabular}{c|c}
% \hline
% \textbf{Parameter} & \textbf{Value} \\
% \hline
% Simulation Sampling Rate  & 2 Sa/sym \\
% PCB Loss (TRX/RX) @ 53\,GHz & 9\,dB \\
% TRX SNR & 30\,dB @200G, 28\,dB @400G \\
% RIN & $-149.5$\,dB/Hz \\
% MZM Bias Voltage & $V_{\pi}/4$ \\
% MZM ER & 4.5\,dB \\
% MZM OMA & 5\,dBm \\
% Optical Channel Attenuation & 3\,dB \\
% RX Butterworth Filter 3\,dB BW & 65\,GHz @200G, 120\,GHz @400G \\
% Thermal noise density & $14~\text{pA}/\sqrt{\text{Hz}}$ \\
% % Shot noise variance & $\sigma_S^2 = 2qI_{\text{PD}}B_{\text{RX}}$\\
% \hline
% \end{tabular}
 
% \end{table}

% \end{wraptable}

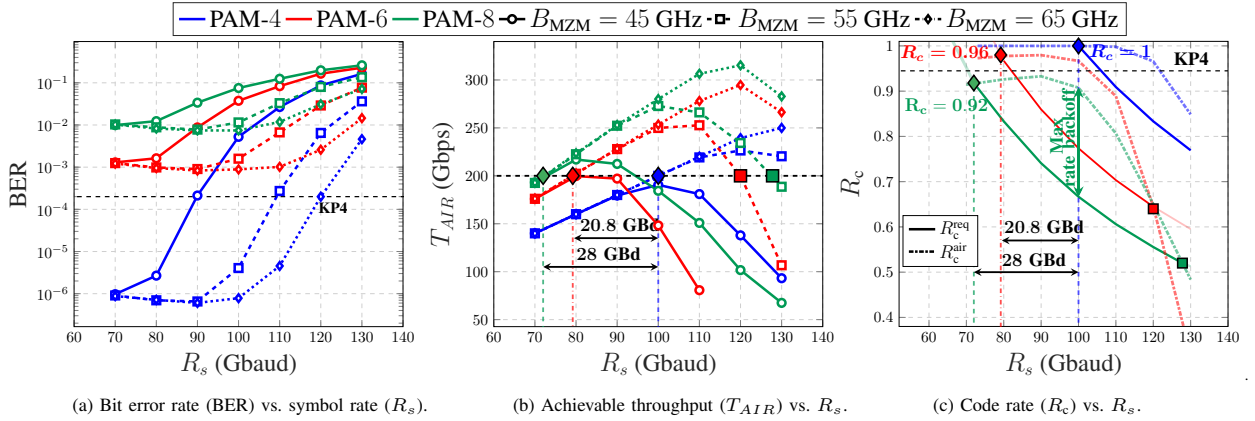
\begin{figure}[t]
    \centering
    \begin{subfigure}[b]{0.31\textwidth}
        %\centering
        \hspace{-.3cm}
        \scalebox{0.58}{\input{Figures/plots/Var_code/200G/BER_OMA_var_coderate_9PCB}}
        \captionsetup{margin={.1cm,-1cm}}\caption{Bit error rate (BER) vs. symbol rate ($R_s$).}
        \label{fig:var_code_rate_BER_RS_PCB_9_200G}
    \end{subfigure}
    \hfill
    \begin{subfigure}[b]{0.32\textwidth}
        %\centering        
        \hspace{-.3cm}
        \scalebox{.58}{\input{Figures/plots/Var_code/200G/AIR_RS_PCB_9_30SNR_200G}}
        \captionsetup{margin={.6cm,-1cm}}
        \caption{Achievable throughput ($T_{AIR}$) vs.  $R_s$.}
        \label{fig:var_code_rate_AIR_9_200G}
    \end{subfigure}
    \hfill
    \begin{subfigure}[b]{0.32\textwidth}
        %\centering
        \hspace{-.4cm}
        \scalebox{.58}{\input{Figures/plots/Var_code/200G/Rc_Rs_PCB9_200G}.}
         \caption{Code rate ($R_{\text{c}}$) vs. $R_s$.}
        \label{fig:var_code_rate_RC_9_200G}
    \end{subfigure}
    % \vspace{-5.5mm}
    \caption{Performance with varying $B_{\text{MZM}}$ ($45$–$65$~GHz) at $30$~dB SNR, $9$~dB PCB loss at $53$~GHz, and RIN = $-149.5$ dB/Hz for the $200$~Gb/s scenario.}% 
    \vspace{-4mm}
\end{figure}

%%%%%%% 200G resutls However, this remains below the $212$~GBd required to support net $400$G transmission with KP4

Fig.~\ref{fig:var_code_rate_BER_RS_PCB_9_200G} shows the BER versus $R_{s}$ for the 200G pre-FEC transmission scenario. Under the fixed-rate KP4 criterion, PAM-4 is the only modulation format that achieves the KP4 FEC threshold, reaching up to $120$~Gbaud with $65$~GHz MZM bandwidth, which is above the $105.83$~GBd required to support a net transmission of $200$G with KP4, but only $90$~Gbaud for $45$~GHz MZM. PAM-6 and PAM-8 fail to meet the KP4 threshold across all tested bandwidths, which under a fixed-rate evaluation would incorrectly suggest these formats are unsuitable.
Across all PAM-$M$ formats, the minimum BER occurs at approx.~$90$~Gbaud, due to the combined TX/RX filtering effects, which become more detrimental as the symbol rate increases. This slight performance dip is due to the CTLE peaking response in Fig.~\ref{fig:all_filters}: the stronger the peaking response due to the higher symbol rate, the steeper the high-frequency roll-off, suppressing the out-of-band noise from all noise contributions except thermal noise, which is introduced after the RX-CTLE, and slightly improves the BER performance. However, beyond a given symbol rate, ISI dominates the performance. At lower symbol rates, maintaining a fixed data rate requires higher modulation orders, which increases the required SNR. Consequently, the BER increases at

\begin{wraptable}[11]{r}{0.42\columnwidth}
%\vspace{-0.5em}
\vspace{-1.4\baselineskip}
\centering
\small
\caption{Simulation Parameters}
\label{tab:fixed_code_rate}
\setlength{\tabcolsep}{1pt}
\renewcommand{\arraystretch}{1}
\begin{tabular}{c|c}
\hline\hline
\textbf{Parameter} & \textbf{Value} \\
\hline\hline
Simulation Sampling Rate & 2 Sa/sym \\\hline
PCB Loss @ 53\,GHz & 9\,dB \\\hline
TRX SNR &\makecell{ 30\,dB @200G,\\ 28\,dB @400G} \\\hline
RIN & $-149.5$\,dB/Hz \\\hline
MZM Bias & $V_{\pi}/4$ \\\hline
MZM Extinction Ratio & 4.5\,dB \\\hline
MZM Opt. Modulation Amplitude & 5\,dBm \\\hline
Optical ch. Attenuation & 3\,dB \\\hline
%RX BW & 65/120\,GHz \\
Thermal noise (RMS)& $14~\text{pA}/\sqrt{\text{Hz}}$ \\\hline
\hline
\end{tabular}
% \vspace{-1em}
\end{wraptable}

\noindent symbol rates below the optimum ($R_s<90$~GBd).

%%%%%%%%%%%%%%%%%%%%%%%%%%%%%%%%%%%%%%%%%%%%%%
The achievable throughput ($T_{AIR}$) is shown in Fig.~\ref{fig:var_code_rate_AIR_9_200G}. In contrast to the fixed-rate KP4 picture, the AIR analysis reveals that $200$~Gb/s can be achieved at $72$, $79.2$, and $100$~Gbaud for PAM-$8$, PAM-$6$, and PAM-$4$, respectively (diamond markers), corresponding to baud-rate reductions of $28$~Gbaud (PAM-8) and $20.8$~Gbaud (PAM-6) relative to PAM-4, achievable by pairing higher-order formats with an appropriately chosen code rate. Square markers achieve the same throughput at higher symbol rates and are of no practical interest. Across the tested MZM 3-dB bandwidths, all formats reach $200$~Gb/s except PAM-4 at $B_{\text{MZM}}=45$~GHz, where ISI is so strong as to prevent the target throughput from being reached.
 
Fig.~\ref{fig:var_code_rate_RC_9_200G} shows the achievable code rate $R_{\text{c}}^{\text{air}}$ supportable by the channel under ideal HD decoding for the $55$~GHz MZM, and the required code rate $R_{\text{c}}^{\text{req}}$ to deliver a net data rate of $200$~Gb/s, both as a function of $R_s$. Any $R_s$-$R_{\text{c}}$ pair satisfying $R_{\text{c}}^{\text{req}} \leq R_{\text{c}}^{\text{air}}$ represents a feasible operating point for $200$~Gb/s transmission, and the two intersections where $R_{\text{c}}^{\text{req}} = R_{\text{c}}^{\text{air}}$ delimit the achievable $R_s$ range. Towards the center of this feasible range, a gap emerges between $R_{\text{c}}^{\text{air}}$ and $R_{\text{c}}^{\text{req}}$, which represents a rate backoff margin that can be used to relax the FEC performance requirements and accommodate the suboptimality of practical FEC schemes. However, exploiting this margin comes at the cost of an increased symbol rate compared to the minimum achievable under a very strong, low-rate FEC. This gap is maximized approximately in the middle of the $R_s$ operating range for each modulation format; for example, in the PAM-8 case, the maximum rate backoff occurs around $R_s=100$~GBd (see vertical green arrow in Fig.~\ref{fig:var_code_rate_AIR_9_200G}). Beyond this point, increasing $R_s$ simultaneously shrinks the rate backoff margin and raises the symbol rate, making operation in this region less attractive. Using KP4 ($R_{\text{c}}^{\text{KP4}} \approx 0.9449$) as a benchmark, the selected operating points (filled markers) correspond to $R_{c}=0.96$ for PAM-6 and $R_{c}=0.92$ for PAM-8, yielding symbol rate reductions of $20.8$~GBd and $28$~GBd relative to PAM-4 while providing a moderate backoff margin below $R_{\text{c}}^{\text{air}}$.
 
% that a modest reduction of the FEC rate by $4\%$ ($R_{c}=0.96$) and $8\%$ ($R_{c}=0.92$) will lead to theoretical reductions of $20.8$ Gbaud and $28$ Gbaud for PAM-6 and PAM-8 with respect to PAM-4, where operating with lower code rates will allow for higher modulation formats. 

\begin{figure}[t]
    % \centering
    \hspace{-.3cm}
    \begin{subfigure}[b]{0.31\textwidth}
        \centering
        \scalebox{0.58}{\input{Figures/plots/Var_code/400G/BER_RS_PCB_9_SNR_28_400G}}
        \captionsetup{margin={-2.3cm,-1cm}}
       \caption{Bit error rate (BER) vs. symbol rate ($R_s$).}
        \label{fig:var_code_rate_BER_RS_PCB_9_400G}
    \end{subfigure}
    \hfill
    \begin{subfigure}[b]{0.31\textwidth}
        % \centering
        \hspace{-.3cm}
        \scalebox{.58}{\input{Figures/plots/Var_code/400G/AIR_RS_PCB_9_SNR_28_400G}}
        \captionsetup{margin={-2.3cm,-1cm}}
        \caption{Achievable throughput ($T_{AIR}$) vs.  $R_s$.}
        \label{fig:var_code_rate_AIR_9_400G}
    \end{subfigure}
    \hfill
    \begin{subfigure}[b]{0.32\textwidth}
        % \centering
        \hspace{-.4cm}
        \scalebox{.58}{\input{Figures/plots/Var_code/400G/Rc_Rs_PCB_9_SNR_28_400G}}
        \caption{Code rate ($R_{\text{c}}$) vs. $R_s$.}
        \label{fig:var_code_rate_RC_9_400G}
    \end{subfigure}
    % \vspace{-5.5mm}
    \caption{Performance with varying $B_{\text{MZM}}$ ($100$–$120$~GHz) at SNR $28$~dB, PCB loss $9$~dB at $53$~GHz, and RIN = $-149.5$ dB/Hz for the $400$~Gb/s scenario.}
     \vspace{-3mm}
\end{figure}
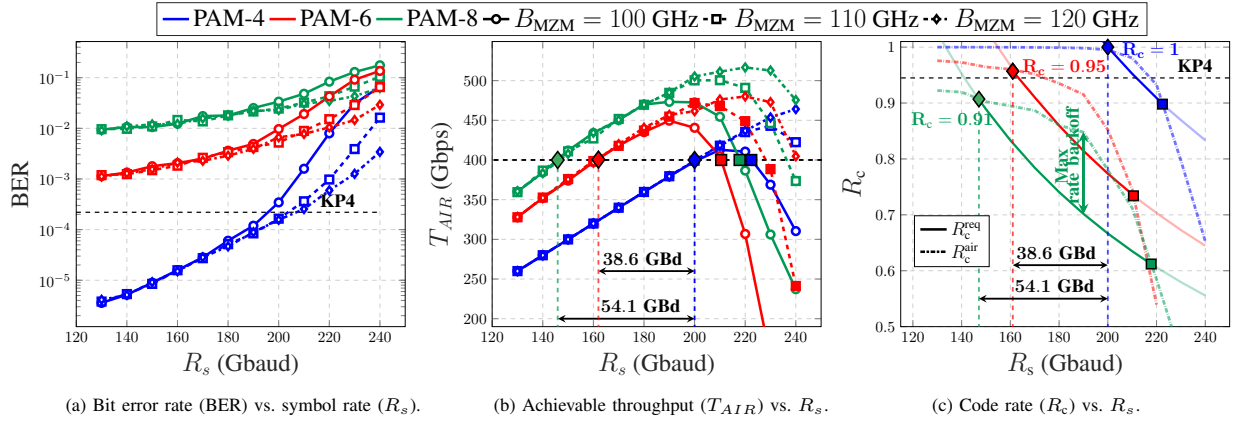

Fig.~\ref{fig:var_code_rate_BER_RS_PCB_9_400G} shows the BER versus $R_s$ for the 400G scenario. Under the fixed-rate KP4 criterion, PAM-4 is the only format that reaches the KP4 FEC threshold, crossing it at approximately $200$ and $205$~GBd for the $110$–$120$~GHz $B_{\text{MZM}}$. However, this remains below the $212$~GBd required to support net $400$G transmission with KP4, and PAM-6 and PAM-8 fail to meet the KP4 threshold entirely.
Figure~\ref{fig:var_code_rate_AIR_9_400G} illustrates the $T_{AIR}$ versus $R_s$. The AIR analysis reveals that $400$~Gb/s can be achieved at symbol rates of $145.9$, $161.4$, and $200$~GBd for PAM-8, PAM-6, and PAM-4, respectively, corresponding to baud rate reductions of $54.1$~GBd (PAM-8) and $38.6$~GBd (PAM-6) relative to PAM-4. All PAM-$M$ formats can reach $400$~Gb/s across all tested MZM bandwidths. Figure~\ref{fig:var_code_rate_RC_9_400G} compares the achievable code rate $R_{\text{c}}^{\text{air}}$ for the $100$~GHz MZM and the required code rate $R_{\text{c}}^{\text{req}}$ as a function of $R_s$. Similar to the $200$~Gb/s case, the intersection between $R_{\text{c}}^{\text{air}}$ and $R_{\text{c}}^{\text{req}}$ delimits the feasible $R_s$ range; outside this range, the target throughput cannot be met even with an ideal FEC. Within the feasible range, a rate backoff margin emerges between $R_{\text{c}}^{\text{air}}$ and $R_{\text{c}}^{\text{req}}$, which relaxes the FEC performance requirements at the cost of a moderately increased symbol rate compared to the minimum achievable under a very strong, low-rate FEC. This margin reaches a maximum at $190$~GBd for PAM-8 (see Fig.~\ref{fig:var_code_rate_AIR_9_400G}); beyond this point, increasing $R_s$ simultaneously reduces the available coding margin and raises the bandwidth requirement. The selected operating point at $R_{\text{c}}=0.91$ for PAM-8 achieves a symbol rate reduction of $54.1$~GBd relative to PAM-4, confirming that the symbol rate advantage of higher-order PAM formats is preserved at $400$~Gb/s, with the rate backoff providing additional margin for practical FEC implementation.
%enables a theoretical $R_s$ reduction up to $54.1$ Gbaud with PAM-8 at $400$ Gb/s.This behavior is consistent at 400~Gb/s, where coding back-off similarly enables higher-order PAM formats to trade excess achievable rate for reduced symbol rate and relaxed bandwidth requirements.

%%%%%%%%%%%$%%%%%%% 400G$
 
% In particular, allowing a coding back-off to $R_c=0.91$ enables PAM-8 to reduce the required symbol rate by up to 54.1~Gbaud relative to PAM-4 at 400~Gb/s. This demonstrates that, also at 400~Gb/s, coding back-off enables higher-order PAM formats to trade excess achievable rate for reduced symbol-rate and bandwidth requirements.
%As a practical benchmark, KP4 FEC threshold lies below PAM-4 and PAM-6 requirements but above PAM-8

%% file: Figures/plots/Var_code/200G/BER_OMA_var_coderate_9PCB.tex
\definecolor{mycolor1}{rgb}{0.00000,0.44700,0.74100}%
\begin{tikzpicture}

\begin{axis}[%
width=7.5cm, 
height=6.5cm,
scale only axis,
xmin=60,
xmax=140,
xlabel style={font=\color{white!15!black}},
xlabel={\Large$R_{s}$ (Gbaud)},
ymode=log,
% ymin=1e-06,
% ymax=1,
yminorticks=true,
ylabel style={font=\color{white!15!black}},
ylabel={\Large BER},
axis background/.style={fill=white},
title style={font=\bfseries},
% title={BER vs OMA for FFE @ 200G},
% axis x line*=bottom,
% axis y line*=left,
grid=major,
grid style={dashed,lightgray!75},
legend style={font=\Large,
  at={(1.75,1.02)},
  anchor=south,
  legend columns=6,
  legend cell align=left,
  align=left,
  draw=white!15!black
},
]
\addlegendimage{color=blue, line width=1.65pt}
\addlegendentry{PAM-$4$}
\addlegendimage{color=red, line width=1.65pt}
\addlegendentry{PAM-$6$}
\addlegendimage{color=ForestGreen, line width=1.65pt}
\addlegendentry{ PAM-$8$}
\addlegendimage{color=black, line width=1.65pt, mark=*, mark options={solid, fill=white,mark size=2.5pt}}
\addlegendentry{ $B_{\text{MZM}}=45$~GHz}
\addlegendimage{color=black, dashed, line width=1.65pt,  mark=square*, mark options={solid, fill=white,mark size=2.5pt}}
\addlegendentry{$B_{\text{MZM}}=55$~GHz}
\addlegendimage{color=black, dashdotted, line width=1.65pt,  mark=diamond*, mark options={solid, fill=white,mark size=2.5pt}}
\addlegendentry{$B_{\text{MZM}}=65$~GHz}

\addplot [color=blue, line width=1.65pt, mark=*, mark options={solid, fill=white,mark size=2.5pt}]
  table[]{%
% 60	0.00199134615384615
70	9.72600077730276e-07
80	2.67933618843683e-06
90	0.000212500000000000
100	0.00524000000000000
110	0.0266725000000000
120	0.0867275000000000
130	0.163055000000000
  };
%\addlegendentry{M = 4; SNRtrx = 30; PCB_Loss = 20; MZM_BW = 45; RIN = -149.5;} 
\addplot [color=blue, dashed, line width=1.65pt, mark=square*, mark options={solid, fill=white,mark size=2.5pt}]
  table[]{%
% 60	0.00203230769230769
70	9.07360406091371e-07
80	6.96687082405345e-07
90	6.53394255874674e-07
100	4.05591572123177e-06
110	0.000272500000000000
120	0.00648500000000000
130	0.0363700000000000
  };
%\addlegendentry{M = 4; SNRtrx = 30; PCB_Loss = 20; MZM_BW = 55; RIN = -149.5;} 
\addplot [color=blue,  line width=1.65pt, dotted,mark=diamond*, mark options={solid, fill=white,mark size=2.5pt}]
  table[]{%
% 60	0.00201865384615385
70	8.76225490196078e-07
80	7.13878597891137e-07
90	6.05053191489362e-07
100	7.82765092274007e-07
110	4.53971119133574e-06
120	0.000199615384615385
130	0.00461500000000000
  };
%\addlegendentry{M = 4; SNRtrx = 30; PCB_Loss = 20; MZM_BW = 65; RIN = -149.5;} 

\addplot [color=red, line width=1.65pt, mark=*, mark options={solid, fill=white,mark size=2.5pt}]
  table[row sep=crcr]{%
70	0.001297\\
80	0.001623\\
90	0.008748\\
100	0.037534\\
110	0.083292\\
120	0.164474\\
130	0.22801\\
};
  
\addplot [color=red, line width=1.65pt,dashed, mark=square*, mark options={solid, fill=white,mark size=2.5pt}]
  table[row sep=crcr]{%
70	0.001248\\
80	0.000984\\
90	0.000903333333333333\\
100	0.001588\\
110	0.006676\\
120	0.029036\\
130	0.076894\\
};

\addplot [color=red, line width=1.65pt,dotted, mark=diamond*, mark options={solid, fill=white,mark size=2.5pt}]
  table[row sep=crcr]{%
70	0.001218\\
80	0.000952666666666667\\
90	0.000832666666666667\\
100	0.000886\\
110	0.00101466666666667\\
120	0.002586\\
130	0.014534\\
};

% \addplot [color=red, line width=1.65pt, mark=*, mark options={solid, fill=white}]
%   table[]{%
% % 60	0.0291480000000000
% 70	0.0268140000000000
% 80	0.0291900000000000
% 90	0.0426780000000000
% 100	0.0633480000000000
% 110	0.0845840000000000
% 120	0.127760000000000
% 130	0.187882000000000
%   };
%\addlegendentry{M = 6; SNRtrx = 30; PCB_Loss = 20; MZM_BW = 45; RIN = -149.5;} 
% \addplot [color=red, line width=1.65pt, dashed,mark=square*, mark options={solid, fill=white}]
%   table[]{%
% % 60	0.0286900000000000
% 70	0.0258700000000000
% 80	0.0232160000000000
% 90	0.0223900000000000
% 100	0.0267520000000000
% 110	0.0407560000000000
% 120	0.0593240000000000
% 130	0.0896480000000000
%   };
%\addlegendentry{M = 6; SNRtrx = 30; PCB_Loss = 20; MZM_BW = 55; RIN = -149.5;} 
% \addplot [color=red, line width=1.65pt,dotted, mark=diamond*, mark options={solid, fill=white}]
%   table[]{%
% % 60	0.0288520000000000
% 70	0.0262800000000000
% 80	0.0230680000000000
% 90	0.0212200000000000
% 100	0.0202580000000000
% 110	0.0241740000000000
% 120	0.0374120000000000
% 130	0.0550040000000000
%   };
%\addlegendentry{M = 6; SNRtrx = 30; PCB_Loss = 20; MZM_BW = 65; RIN = -149.5;} 

\addplot [color=ForestGreen, line width=1.65pt, mark=*, mark options={solid, fill=white,mark size=2.5pt}]
  table[]{%
% 60	0.0611833333333333
70	0.00992833333333333
80	0.0123383333333333
90	0.0338233333333333
100	0.0752883333333333
110	0.124721666666667
120	0.197850000000000
130	0.260185000000000
  };
%\addlegendentry{M = 8; SNRtrx = 30; PCB_Loss = 20; MZM_BW = 45; RIN = -149.5;} 
\addplot [color=ForestGreen, dashed,line width=1.65pt, mark=square*, mark options={solid, fill=white,mark size=2.5pt}]
  table[]{%
% 60	0.0609983333333333
70	0.0103116666666667
80	0.00881333333333333
90	0.00778000000000000
100	0.0114983333333333
110	0.0329350000000000
120	0.0798883333333333
130	0.135743333333333
  };
%\addlegendentry{M = 8; SNRtrx = 30; PCB_Loss = 20; MZM_BW = 45; RIN = -149.5;} 
\addplot [color=ForestGreen,dotted, line width=1.65pt, mark=diamond*, mark options={solid, fill=white,mark size=2.5pt}]
  table[]{%
% 60	0.0606250000000000
70	0.0102000000000000
80	0.00823833333333333
90	0.00742666666666667
100	0.00742000000000000
110	0.0118250000000000
120	0.0309566666666667
130	0.0716866666666667
  };
%\addlegendentry{M = 8; SNRtrx = 30; PCB_Loss = 20; MZM_BW = 45; RIN = -149.5;} 

\addplot [black, thick, dashed, domain=0:150] {2e-4}
    node[midway, below=-0.2pt,xshift=4.5cm] {\textbf{KP4}};

\end{axis}

\end{tikzpicture}%

%% file: Figures/plots/Var_code/200G/AIR_RS_PCB_9_30SNR_200G.tex
\definecolor{mycolor}{RGB}{255,237,160}
\definecolor{myblue}{RGB}{4,90,141}
\definecolor{mygreen}{RGB}{65,171,93}

\begin{tikzpicture}

\begin{axis}[%
width=7.5cm, 
height=6.5cm,
scale only axis,
xmin=60,
xmax=140,
xlabel style={font=\color{white!15!black}},
xlabel={\Large$R_s$ (Gbaud)},
% ymode=log,
% ymin=1e-06,
% ymax=1,
yminorticks=true,
ylabel style={font=\color{white!15!black}},
ylabel={\Large $T_{AIR}$ (Gbps)},
axis background/.style={fill=white},
title style={font=\bfseries},
% title={BER vs OMA for FFE @ 200G},
% axis x line*=bottom,
% axis y line*=left,
grid=major,
grid style={dashed,lightgray!75},
legend style={font=\small, legend cell align=left, at={(0.02,0.77)}, anchor=south west}, 
]
% \addlegendimage{color=blue, line width=1.65pt}
% \addlegendentry{PAM4}
% \addlegendimage{color=red, line width=1.65pt}
% \addlegendentry{PAM6}
% \addlegendimage{color=ForestGreen, line width=1.65pt}
% \addlegendentry{PAM8}
% \addlegendimage{color=black, line width=1.65pt, mark=*, mark options={solid, fill=white}}
% \addlegendentry{45 GHz}
% \addlegendimage{color=black, dashed, line width=1.65pt, mark=square*, mark options={solid, fill=white}}
% \addlegendentry{55 GHz}
% \addlegendimage{color=black, dotted, line width=1.65pt, mark=diamond*, mark options={solid, fill=white}}
% \addlegendentry{65 GHz}
% \addlegendimage{color=black, line width=1.65pt, mark=triangle*}
% \addlegendentry{9 dB}
% \addlegendimage{color=black, line width=1.65pt, mark=*}
% \addlegendentry{20 dB}

\addplot [color=blue, line width=1.65pt, mark=*, mark options={solid, fill=white, mark size=2.5pt}]
  table[]{%
% 60	117.511622314957
70	139.997084136990
80	159.991446535996
90	179.478163233464
100	190.552147917131
110	180.967660232267
120	137.891696246927
130	93.1922555235235
};
%\addlegendentry{M = 4}
\addplot [color=blue, dashed, line width=1.65pt, mark=square*, mark options={solid, fill=white, mark size=2.5pt}]
  table[]{%
% 60	117.467608006594
70	139.997267001287
80	159.997559289970
90	179.997413942150
100	199.984300172413
110	219.203626921401
120	226.448915983747
130	220.397631484312
};
%\addlegendentry{M = 4}
\addplot [color=blue, dotted, line width=1.65pt, mark=diamond*, mark options={solid, fill=white, mark size=2.5pt}]
  table[]{%
% 60	117.482263385599
70	139.997354601253
80	159.997503079685
90	179.997593193233
100	199.996598476801
110	219.980832589573
120	239.342081721111
130	249.962342450393
};
%\addlegendentry{M = 4}

\addplot [color=blue,line width=1pt  ,mark=diamond*, draw=black,mark options={solid, fill=blue, mark size=5.5pt}]
  table[row sep=crcr]{%
100 200\\
};

% \addplot [color=red, line width=1.65pt, mark=*, mark options={solid, fill=white}]
%   table[]{%
% 60	125.613519090778
% 70	148.710969558481
% 80	167.440772176352
% 90	173.452300156941
% 100	170.453902258761
% 110	165.452691980656
% 120	139.196254961026
% 130	101.815058919459
% };
% %\addlegendentry{M = 6}
% \addplot [color=red, dashed, line width=1.65pt, mark=square*, mark options={solid, fill=white}]
%   table[]{%
% 60	125.973631071866
% 70	149.600580022735
% 80	173.888292690400
% 90	196.666144927356
% 100	212.527314930266
% 110	214.468516457513
% 120	209.457778550787
% 130	189.766767514793
% };
% %\addlegendentry{M = 6}
% \addplot [color=red, dotted, line width=1.65pt, mark=diamond*, mark options={solid, fill=white}]
%   table[]{%
% 60	125.846064035805
% 70	149.213087477616
% 80	174.053551565632
% 90	198.159840982320
% 100	221.560574415126
% 110	237.635019621402
% 120	238.758158755931
% 130	232.783575823329
% };
% %\addlegendentry{M = 6}

\addplot [color=red, line width=1.65pt, mark=*, mark options={solid, fill=white, mark size=2.5pt}]
  table[row sep=crcr]{%
70	175.901693697312\\
80	199.736018903595\\
90	197.107382684695\\
100	148.253320056806\\
110	80.6998207146119\\
% 120	11.2640812822428\\
% 130	0.169863767489214\\
};
% \addlegendentry{M=6, MZM\_BW=45GHz}

\addplot [color=red, line width=1.65pt,dashed, mark=square*, mark options={solid, fill=white, mark size=2.5pt}]
  table[row sep=crcr]{%
70	176.036156294954\\
80	202.167527795584\\
90	227.88784454349\\
100	249.933411299847\\
110	252.74906408027\\
120	200.179962646656\\
130	106.690905023065\\
};
% \addlegendentry{M=6, MZM\_BW=55GHz}

\addplot [color=red, line width=1.65pt, dotted,mark=diamond*, mark options={solid, fill=white, mark size=2.5pt}]
  table[row sep=crcr]{%
70	176.148124375084\\
80	202.461841582373\\
90	228.29206716582\\
100	253.449942826409\\
110	278.060048452466\\
120	294.76031385002\\
130	266.336321354337\\
};
% \addlegendentry{M=6, MZM\_BW=65GHz}

\addplot [color=red,line width=1pt ,mark=diamond*,opacity=1,draw=black, mark options={solid, fill=red, mark size=5.5pt}]
  table[row sep=crcr]{%
79.2 200\\
};

\addplot [color=ForestGreen, line width=1.65pt, mark=*, mark options={solid, fill=white, mark size=2.5pt}]
  table[]{%
% 60	120.217588175252
70	193.133292054738
80	216.978249372352
90	212.431312471647
100	184.393277327294
110	150.881758006270
120	101.661408586634
130	67.4600894290013
};
%\addlegendentry{M = 8}
\addplot [color=ForestGreen, dashed, line width=1.65pt, mark=square*, mark options={solid, fill=white, mark size=2.5pt}]
  table[]{%
% 60	120.348855387206
70	192.601021329640
80	222.523339117675
90	252.264460951448
100	272.828977701749
110	266.163
120	234.287
130	188.612
};
%\addlegendentry{M = 8}
\addplot [color=ForestGreen, dotted, line width=1.65pt, mark=diamond*, mark options={solid, fill=white, mark size=2.5pt}]
  table[]{%
% 60	120.614227966000
70	192.755616537040
80	223.470276350976
90	252.934957992127
100	280.052968148727
110	306.421394236431
120	315.299866711253
130	282.847199148401
};
%\addlegendentry{M = 8}
\addplot [color=mygreen,line width=1pt ,mark=diamond*,draw=black,mark options={solid, fill=mygreen,mark size=5.5pt}]
  table[row sep=crcr]{%
72 200\\
};

\addplot [color=red,line width=1pt  ,mark=square*, draw=black,mark options={solid, fill=red, mark size=3.9pt}]
  table[row sep=crcr]{%
120 200\\
};

\addplot [color=ForestGreen,line width=1pt  ,mark=square*, draw=black,mark options={solid, fill=ForestGreen, mark size=3.9pt}]
  table[row sep=crcr]{%
127.8 200\\
};

\draw[dashed,very thick,ForestGreen,opacity=0.7] (axis cs:72,0) -- (axis cs:72,200);

\draw[dashdotted,very thick,red,opacity=0.7] (axis cs:79.2,0) -- (axis cs:79.2,200);

\draw[dashed,very thick, blue,opacity=0.7] (axis cs:100,0) -- (axis cs:100,200);

% \path[name path=axis1] (axis cs:25,200) -- (axis cs:270,200);
% \path[name path=axis2]  (axis cs:25,0) -- (axis cs:270,0);

% \addplot[gray, opacity=0.2] fill between[of=axis1 and axis2];

\addplot [black, very thick, dashed, domain=0:300] {200}
    node[midway, below=-0.2pt,xshift=-2.5cm] {};

\draw[<->,style={draw,>=stealth,rounded corners},very thick,black] (axis cs:72,105) -- (axis cs: 100,105) node[midway,above=0pt, yshift=+.001cm, xshift=.18cm] {\large $\mathbf{28}$ \textbf{GBd}};

\draw[<->,style={draw,>=stealth,rounded corners},very thick,black] (axis cs:79.5,135) -- (axis cs: 100,135) node[midway,above=0.5pt, yshift=+.001cm, xshift=+.09cm] {\large $\mathbf{20.8}$ \textbf{GBd}};

% \path[<->,very thick, draw,>=latex] (axis cs:72,200)--++(3pt,0pt) node[midway,below]{28 Gbaud}--(axis cs:100,200);

\end{axis}

\end{tikzpicture}%

%% file: Figures/plots/Var_code/200G/Rc_Rs_PCB9_200G.tex
\definecolor{mycolor}{RGB}{255,237,160}
\definecolor{myblue}{RGB}{65,182,196}
\definecolor{mygreen}{RGB}{65,171,93}

\begin{tikzpicture}

\begin{axis}[%
width=7.5cm, 
height=6.5cm,
scale only axis,
xmin=52,
xmax=140,
xlabel style={font=\color{white!15!black}},
xlabel={\Large$R_s$ (Gbaud)},
% ymode=log,
ymin=0.38,
ymax=1.01,
yminorticks=true,
ylabel style={font=\color{white!15!black}},
ylabel={\Large$R_{\text{c}}$},
axis background/.style={fill=white},
title style={font=\bfseries},
% title={BER vs OMA for FFE @ 200G},
% axis x line*=bottom,
% axis y line*=left,
grid=major,
grid style={dashed,lightgray!75},
legend style={font=\large, legend cell align=left, at={(0.24,0.39)}, anchor=north east}, 
]

% \addlegendimage{color=blue, line width=1.65pt}
% \addlegendentry{PAM4}
% \addlegendimage{color=red, line width=1.65pt}
% \addlegendentry{PAM6}
% \addlegendimage{color=ForestGreen, line width=1.65pt}
% \addlegendentry{PAM8}
\addlegendimage{color=black, line width=1.65pt}
\addlegendentry{ $R_{\text{c}}^{\text{req}}$}
\addlegendimage{color=black, line width=1.65pt,densely dashdotted}
\addlegendentry{ $R_{\text{c}}^\text{air}$}
% \addlegendimage{color=black, line width=1.65pt, mark=diamond*, mark options={solid, fill=white}}
% \addlegendentry{90 GHz}
% \addlegendimage{color=black, line width=1.65pt,dashed, mark=square*, mark options={solid, fill=white}}
% \addlegendentry{100 GHz}
% \addlegendimage{color=black, line width=1.65pt, dashdotted,mark=diamond*, mark options={solid, fill=white}}
% \addlegendentry{110 GHz}
% \addlegendimage{color=black, line width=1.65pt, dotted, mark=triangle* , mark options={solid, fill=white}}
% \addlegendentry{120 GHz}
% \addlegendimage{color=blue, line width=1.5pt}
% \addlegendentry{PAM4}
% \addlegendimage{color=red, line width=1.5pt}
% \addlegendentry{PAM6}
% \addlegendimage{color=ForestGreen, line width=1.5pt}
% \addlegendentry{PAM8}
% \addlegendimage{color=black, line width=1.5pt, mark=*}
% \addlegendentry{PCB9}
% \addlegendimage{color=black, line width=1.5pt, mark=square*}
% \addlegendentry{55 GHz}
% \addlegendimage{color=black, dashed, line width=1.5pt, mark=diamond*}
% \addlegendentry{PCB20}

\addplot [color=blue, line width=1.5pt]
  table[]{%
60	1.66666666666667
70	1.42857142857143
80	1.25000000000000
90	1.11111111111111
100	1
110	0.909090909090909
120	0.833333333333333
130	0.769230769230769
};
%\addlegendentry{M = 4}
% \addplot [color=red, line width=1.5pt, mark=*, mark options={solid, fill=white, red}]

% \addplot [color=red,dotted, line width=2pt]
%   table[]{%
% 60	1.28950935744847
% 70	1.10529373495583
% 80	0.967132018086354
% 80.1 0.965
% };
% \addplot [color=red, line width=1.5pt]
%   table[]{%
% 80.1 0.965
% 90	0.859672904965648
% 100	0.773705614469083
% 110	0.703368740426439
% 120	0.644754678724236
% 130	0.595158164976218
% };

\addplot [color=red, opacity=0.25, line width=1.2pt]
  table[row sep=crcr]{%
60	1.28950935744847
70	1.10529373495583\\
79.2 0.98\\
};

\addplot [color=red, line width=1.2pt]
  table[row sep=crcr]{%
80	0.967132018086354\\
90	0.859672904965648\\
100	0.773705614469083\\
110	0.703368740426439\\
120	0.644754678724236\\
% 130	0.595158164976218\\
};

\addplot [color=red, line width=1.2pt,opacity=0.25]
  table[row sep=crcr]{%
120	0.644754678724236\\
130	0.595158164976218\\
};

% \addlegendentry{M=6, MZM\_BW=55GHz (required)}

% \addlegendentry{M = 6}
% \addplot [color=ForestGreen, line width=1.5pt, mark=*, mark options={solid, fill=white, ForestGreen}]

\addplot [color=ForestGreen,opacity=0.25, line width=2pt]
  table[]{%
60	1.11111111111111
70	0.952380952380952
72 0.918
};

\addplot [color=ForestGreen, line width=1.5pt]
  table[]{%
72 0.918
80	0.833333333333333
90	0.740740740740741
100	0.666666666666667
110	0.606060606060606
120	0.555555555555556
127.8 0.52
};

\addplot [color=ForestGreen, line width=1.5pt,opacity=0.25]
  table[]{%
127.8 0.52
130	0.512820512820513
};

%\addlegendentry{M = 8}

\addplot+[only marks, mark=star*, mark options={fill=white,draw=black}, mark size=3pt]
coordinates {(106.25, 0.9449)};

% \draw[dashed,very thick,ForestGreen,opacity=1.1] (axis cs:60,0.9252) -- (axis cs:72,0.9252) node[midway,below=0.2pt, yshift=-.001cm, xshift=-.09cm] {};

\addplot[
  only marks,
  mark=star*,
  mark size=7pt,
  draw=ForestGreen,
  fill=ForestGreen
] coordinates {(72,0.9252)};
\node[anchor=west, text=ForestGreen,yshift=-0.6cm, xshift=-1.7cm,rotate=0]
  at (axis cs:72,0.9252) {\large $\mathbf{R_{c} = 0.92}$};

\addplot [color=mygreen,line width=1pt  ,mark=diamond*, draw=black,mark options={solid, fill=mygreen, mark size=5pt}]
  table[row sep=crcr]{%
72 0.918\\
};

% \draw[dashed,very thick,ForestGreen,opacity=0.25] (axis cs:72,0) -- (axis cs:72,0.9252);
% \draw[dashdotted,very thick,red,opacity=0.25] (axis cs:79.2,0) -- (axis cs:79.2,0.98);
% \draw[dashdotted,very thick,blue,opacity=0.25] (axis cs:100,0) -- (axis cs:100,1);

% \draw[dashdotted,very thick,red,opacity=1.1] (axis cs:60,0.965) -- (axis cs:80.1,0.965) node[midway,above=0.2pt, yshift=-.001cm, xshift=+1.0cm] {};

% Second point (red)
\addplot[
  only marks,
  mark=star*,
  mark size=2.5pt,
  draw=red,
  fill=red
] coordinates {(80.1,0.965)};
\node[anchor=west, text=red,yshift=0.1cm, xshift=-.39cm]
  at (axis cs:55.5,0.975) {\large $\boldsymbol{R_{c} = 0.96}$};

\addplot [color=red,line width=1pt  ,mark=diamond*, draw=black,mark options={solid, fill=red, mark size=5pt}]
  table[row sep=crcr]{%
79.2 0.98\\
};

\addplot [color=blue,line width=1pt  ,mark=diamond*, draw=black,mark options={solid, fill=blue, mark size=5pt}]
  table[row sep=crcr]{%
100 1\\
};

\addplot[
  only marks,
  mark=star*,
  mark size=2.5pt,
  draw=blue,
  fill=blue
] coordinates {(100,0.99)};
\node[anchor=west, text=blue,yshift=-0.25cm, xshift=.3cm]
  at (axis cs:98,1.01) {\large $\boldsymbol{R_{c} = 1}$};

\addplot [color=blue, line width=1.8pt,densely dashdotted,opacity=0.55]
  table[row sep=crcr]{%
73	0.999974593687814\\
80	0.999979907103381\\
90	0.999980128841678\\
100	0.999924303520823\\
110	0.998003938888226\\
120	0.965794563779048\\
130	0.8485825878679\\
};
% \addlegendentry{M=4, MZM\_BW=55GHz (ideal)}

\addplot [color=red, line width=1.8pt,densely dashdotted, opacity=0.55]
  table[row sep=crcr]{%
73	0.972858303392592\\
80	0.977613445742359\\
90	0.979545026625309\\
100	0.96687441783051\\
110	0.888878954230504\\
120	0.64533483751637\\
130	0.317489816265897\\
};

% \addlegendentry{M=6, MZM\_BW=55GHz (ideal)}

\addplot [color=ForestGreen, line width=1.8pt,densely dashdotted, opacity=0.55]
  table[row sep=crcr]{%
70	0.914177010767547\\
80	0.926035995146053\\
90	0.932801636906151\\
100	0.907614351378107\\
110	0.806553505508307\\
120	0.650798137868048\\
130	0.483621097938648\\
};

%%%%%%%%%%%%%%%%%%%%%%%%%%%%%%%%%%%%%%%%%%%%%%%%%%%%%%%%%%%%%%%%%%%%%%%%%%%%%%%%%%%%%%%%%%%%%%%%%%%%%%%%%%%%%%%%%%%%%%%%%%%   crossing points squares%%%%%%%%%%%%%%%%%%%%%%%%%%%%%%%%%%%%%%%%% %%%%%%%%%%%%%%%%%%%%%%%%%%%%%%%%%%%%%%%%% %%%%%%%%%%%%%%%%%%%%%%%%%%%%%%%%%%%%%%%%% 

\addplot [color=red,line width=1pt  ,mark=square*, draw=black,mark options={solid, fill=red, mark size=3.1pt}]
  table[row sep=crcr]{%
120 0.64\\
};

\addplot[
  only marks,
  mark=star*,
  mark size=2.5pt,
  draw=red,
  fill=red
] coordinates {(100,0.99)};
% \node[anchor=west, text=red,yshift=0.01cm, xshift=.1cm]
%   at (axis cs:120,0.66) {\normalsize$\mathbf{r \approx 0.64}$};

\addplot [color=ForestGreen,line width=1pt  ,mark=square*, draw=black,mark options={solid, fill=ForestGreen, mark size=3.1pt}]
  table[row sep=crcr]{%
127.8 0.52\\
};

\addplot[
  only marks,
  mark=star*,
  mark size=2.5pt,
  draw=ForestGreen,
  fill=ForestGreen
] coordinates {(100,0.99)};
% \node[anchor=west, text=ForestGreen,yshift=-0.25cm, xshift=-1.8cm]
%   at (axis cs:127.8,0.54) {\normalsize$\mathbf{r \approx 0.52}$};

\addplot [black, thick, dashed, domain=50:140] {0.9449}
    node[midway, above=0.05pt,xshift=3cm] {\large \textbf{KP4}};

\draw[dashed,very thick,ForestGreen,opacity=0.7] (axis cs:72,0) -- (axis cs:72,0.92);

\draw[dashdotted,very thick,red,opacity=0.7] (axis cs:79.2,0) -- (axis cs:79.2,0.96);

\draw[dashed,very thick, blue,opacity=0.7] (axis cs:100,0) -- (axis cs:100,1);

\draw[<->,style={draw,>=stealth,rounded corners},very thick,black] (axis cs:72,0.5) -- (axis cs: 100,0.5) node[midway,above=0pt, yshift=+.001cm, xshift=.18cm] {\large $\mathbf{28}$ \textbf{GBd}};

\draw[<->,style={draw,>=stealth,rounded corners},very thick,black] (axis cs:79.5,0.57) -- (axis cs: 100,0.57) node[midway,above=0.5pt, yshift=+.001cm, xshift=+.09cm] {\large $\mathbf{20.8}$ \textbf{GBd}};

%%%% MAX Back Off
\draw[<->,style={draw,>=stealth,rounded corners},
very thick,ForestGreen]
(axis cs:100,0.666666666666667) --
(axis cs:100,0.907614351378107)
node[midway,
above=0pt,
yshift=+0.151cm,
xshift=-.21cm,
rotate=90,
align=center]
{\large \textbf{Max}};

\draw[<->,style={draw,>=stealth,rounded corners},
very thick,line width=1.9pt,ForestGreen]
(axis cs:100,0.666666666666667) --
(axis cs:100,0.907614351378107)
node[midway,
above=0pt,
yshift=+0.0901cm,
xshift=0.07cm,
rotate=90,
align=center]
{\large \textbf{rate backoff}};

% 100	0.666666666666667
% 100	0.907614351378107

\end{axis}

\end{tikzpicture}%x

%% file: Figures/plots/Var_code/400G/BER_RS_PCB_9_SNR_28_400G.tex
\begin{tikzpicture}

\begin{axis}[%
width=7.5cm, 
height=6.5cm,
scale only axis,
xmin=120,
xmax=250,
xlabel style={font=\color{white!15!black}},
xlabel={\Large $R_s$ (Gbaud)},
ymode=log,
% ymin=1e-06,
% ymax=1,
yminorticks=true,
ylabel style={font=\color{white!15!black}},
ylabel={\Large BER},
axis background/.style={fill=white},
title style={font=\bfseries},
% title={BER vs OMA for FFE @ 200G},
% axis x line*=bottom,
% axis y line*=left,
grid=major,
grid style={dashed,lightgray!75},
legend style={font=\Large,
  at={(1.75,1.02)},
  anchor=south,
  legend columns=6,
  legend cell align=left,
  align=left,
  draw=white!15!black
},
]

\addlegendimage{color=blue, line width=1.65pt}
\addlegendentry{PAM-4}
\addlegendimage{color=red, line width=1.65pt}
\addlegendentry{PAM-6}
\addlegendimage{color=ForestGreen, line width=1.65pt}
\addlegendentry{ PAM-8}
\addlegendimage{color=black, line width=1.65pt, mark=*, mark options={solid, fill=white, mark size=2.5pt}}
\addlegendentry{ $B_{\text{MZM}}=100$ GHz}
\addlegendimage{color=black, line width=1.65pt,dashed, mark=square*, mark options={solid, fill=white, mark size=2.5pt}}
\addlegendentry{ $B_{\text{MZM}}=110$ GHz}
\addlegendimage{color=black, line width=1.65pt, dashdotted,mark=diamond*, mark options={solid, fill=white, mark size=2.5pt}}
\addlegendentry{$B_{\text{MZM}}=120$ GHz}
% \addlegendimage{color=black, line width=1.65pt, dotted, mark=triangle* , mark options={solid, fill=white}}
% \addlegendentry{120 GHz}

% \addplot [color=blue, line width=1.65pt, mark=*, mark options={solid, fill=white}]
%   table[row sep=crcr]{%
% 130	3.85593220338983e-06\\
% 140	5.71022727272727e-06\\
% 150	9.00179856115108e-06\\
% 160	1.76232394366197e-05\\
% 170	4.89423076923077e-05\\
% 180	0.00016203125\\
% 190	0.000960833333333333\\
% 200	0.006065\\
% 210	0.027125\\
% 220	0.063665\\
% 230	0.105635\\
% 240	0.143195\\
% };
%\addlegendentry{M=4, MZM\_BW=90GHz}

% \addplot [color=ForestGreen, line width=1.65pt, mark=*, mark options={solid, fill=white}]
%   table[row sep=crcr]{%
% 130	0.00956666666666667\\
% 140	0.00934833333333333\\
% 150	0.0118233333333333\\
% 160	0.0132616666666667\\
% 170	0.019205\\
% 180	0.026565\\
% 190	0.0456166666666667\\
% 200	0.07652\\
% 210	0.122528333333333\\
% 220	0.168575\\
% 230	0.213241666666667\\
% 240	0.250101666666667\\
% };
%\addlegendentry{M=8, MZM\_BW=90GHz}

\addplot [color=blue, line width=1.65pt, mark=*, mark options={solid, fill=white, mark size=2.5pt}]
  table[row sep=crcr]{%
130	3.575e-06\\
140	5.18115942028985e-06\\
150	8.9375e-06\\
160	1.58333333333333e-05\\
170	2.82584269662921e-05\\
180	6.03571428571429e-05\\
190	0.000119404761904762\\
200	0.0003446875\\
210	0.0015925\\
220	0.0079575\\
230	0.0307375\\
240	0.0666625\\
};
%\addlegendentry{M=4, MZM\_BW=100GHz}

\addplot [color=ForestGreen, line width=1.65pt, mark=*, mark options={solid, fill=white, mark size=2.5pt}]
  table[row sep=crcr]{%
130	0.009385\\
140	0.0106533333333333\\
150	0.0105733333333333\\
160	0.0123183333333333\\
170	0.0174583333333333\\
180	0.017955\\
190	0.0251966666666667\\
200	0.0336416666666667\\
210	0.0481716666666667\\
220	0.0834233333333333\\
230	0.129615\\
240	0.175695\\
};
%\addlegendentry{M=8, MZM\_BW=100GHz}

\addplot [color=blue, line width=1.65pt,dashed, mark=square*, mark options={solid, fill=white, mark size=2.5pt}]
  table[row sep=crcr]{%
130	3.79742033383915e-06\\
140	5.33510638297872e-06\\
150	8.48484848484848e-06\\
160	1.57547169811321e-05\\
170	2.72826086956522e-05\\
180	5.16326530612245e-05\\
190	8.43333333333333e-05\\
200	0.00016453125\\
210	0.000362857142857143\\
220	0.000975833333333333\\
230	0.0038975\\
240	0.0162625\\
};
%\addlegendentry{M=4, MZM\_BW=110GHz}

\addplot [color=ForestGreen, line width=1.65pt, dashed, mark=square*, mark options={solid, fill=white, mark size=2.5pt}]
  table[row sep=crcr]{%
130	0.00962666666666667\\
140	0.00978166666666667\\
150	0.01093\\
160	0.0146416666666667\\
170	0.013705\\
180	0.01806\\
190	0.0215083333333333\\
200	0.0245283333333333\\
210	0.03225\\
220	0.043075\\
230	0.0665666666666667\\
240	0.103896666666667\\
};
%\addlegendentry{M=8, MZM\_BW=110GHz}

\addplot [color=blue, line width=1.65pt, dashdotted, mark=diamond*, mark options={solid, fill=white, mark size=2.5pt}]
  table[row sep=crcr]{%
130	4.11065573770492e-06\\
140	5.13860369609856e-06\\
150	9.1514598540146e-06\\
160	1.50903614457831e-05\\
170	2.83988764044944e-05\\
180	4.6712962962963e-05\\
190	8.87931034482759e-05\\
200	0.000156323529411765\\
210	0.000255\\
220	0.000594\\
230	0.001265\\
240	0.0034275\\
};
%\addlegendentry{M=4, MZM\_BW=120GHz}

\addplot [color=ForestGreen, line width=1.65pt,dashdotted, mark=diamond*, mark options={solid, fill=white, mark size=2.5pt}]
  table[row sep=crcr]{%
130	0.00916166666666667\\
140	0.0109116666666667\\
150	0.011975\\
160	0.0122283333333333\\
170	0.0157016666666667\\
180	0.0180116666666667\\
190	0.0220216666666667\\
200	0.0229883333333333\\
210	0.0307366666666667\\
220	0.034665\\
230	0.0432183333333333\\
240	0.0630833333333333\\
};
%\addlegendentry{M=8, MZM\_BW=120GHz}

%%% PAM-6
% \addplot [color=red, line width=1.65pt, mark=*, mark options={solid, fill=white}]
%   table[row sep=crcr]{%
% 130	0.001147\\
% 140	0.001408\\
% 150	0.001552\\
% 160	0.002174\\
% 170	0.003148\\
% 180	0.007028\\
% 190	0.016286\\
% 200	0.037772\\
% 210	0.081164\\
% 220	0.13388\\
% 230	0.178568\\
% 240	0.213452\\
% };
% \addlegendentry{M=6, MZM\_BW=90GHz}

\addplot [color=red, line width=1.65pt, mark=*, mark options={solid, fill=white, mark size=2.5pt}]
  table[row sep=crcr]{%
130	0.001141\\
140	0.001341\\
150	0.001788\\
160	0.002055\\
170	0.002514\\
180	0.003618\\
190	0.00492\\
200	0.009682\\
210	0.019138\\
220	0.04263\\
230	0.091428\\
240	0.136182\\
};
% \addlegendentry{M=6, MZM\_BW=100GHz}

\addplot [color=red, line width=1.65pt, mark=square*,dashed, mark options={solid, fill=white, mark size=2.5pt}]
  table[row sep=crcr]{%
130	0.001191\\
140	0.001245\\
150	0.00149\\
160	0.001831\\
170	0.002622\\
180	0.003384\\
190	0.004276\\
200	0.005226\\
210	0.0089\\
220	0.015172\\
230	0.028812\\
240	0.065214\\
};
% \addlegendentry{M=6, MZM\_BW=110GHz}
\addplot [color=red, line width=1.65pt, mark=diamond*, dashdotted ,mark options={solid, fill=white, mark size=2.5pt}]
  table[row sep=crcr]{%
130	0.001118\\
140	0.001288\\
150	0.001561\\
160	0.002108\\
170	0.002286\\
180	0.002878\\
190	0.003934\\
200	0.006658\\
210	0.007752\\
220	0.010456\\
230	0.014588\\
240	0.029156\\
};
% \addlegendentry{M=6, MZM\_BW=120GHz}

\addplot [black, thick, dashed, domain=120:240] {2.2e-4}
    node[midway, above=-0.2pt,xshift=2.5cm] {\large \textbf{KP4}};

\end{axis}
\end{tikzpicture}%

%% file: Figures/plots/Var_code/400G/AIR_RS_PCB_9_SNR_28_400G.tex
\definecolor{mycolor}{RGB}{255,237,160}
\definecolor{myblue}{RGB}{127,205,187}
\definecolor{mygreen}{RGB}{65,171,93}

\begin{tikzpicture}

\begin{axis}[%
width=7.5cm, 
height=6.5cm, 
scale only axis,
xmin=120,
xmax=250,
ymin=190,
xlabel style={font=\color{white!15!black}},
xlabel={\Large $R_s$ (Gbaud)},
% ymode=log,
% ymin=1e-06,
% ymax=1,
yminorticks=true,
ylabel style={font=\color{white!15!black}},
ylabel={\Large $T_{AIR}$ (Gbps)},
axis background/.style={fill=white},
title style={font=\bfseries},
% title={BER vs OMA for FFE @ 200G},
% axis x line*=bottom,
% axis y line*=left,
grid=major,
grid style={dashed,lightgray!75},
legend style={font=\scriptsize, legend cell align=left, at={(0.01,0.73)}, anchor=south west}, 
]
% \addlegendimage{color=black, line width=1.65pt, mark=*, mark options={solid, fill=white}}
% \addlegendentry{90 GHz}
% \addlegendimage{color=black, line width=1.65pt,dashed, mark=square*, mark options={solid, fill=white}}
% \addlegendentry{100 GHz}
% \addlegendimage{color=black, line width=1.65pt, dashdotted,mark=diamond*, mark options={solid, fill=white}}
% \addlegendentry{110 GHz}
% \addlegendimage{color=black, line width=1.65pt, dotted, mark=triangle* , mark options={solid, fill=white}}
% \addlegendentry{120 GHz}

% \addplot [color=blue, line width=1.65pt, mark=*, mark options={solid, fill=white}]
%   table[row sep=crcr]{%
% 130	259.980523427718\\
% 140	279.969844294509\\
% 150	299.950839259318\\
% 160	319.902805191509\\
% 170	339.737727269102\\
% 180	359.181378161326\\
% 190	375.813781182674\\
% 200	378.642495910258\\
% 210	344.499826920836\\
% 220	289.597008199018\\
% 230	236.160434290843\\
% 240	195.578181779871\\
% };
%\addlegendentry{M=4, MZM\_BW=90GHz}

% \addplot [color=ForestGreen, line width=1.65pt, mark=*, mark options={solid, fill=white}]
%   table[row sep=crcr]{%
% 130	359.616452292988\\
% 140	387.894622758232\\
% 150	408.306689949517\\
% 160	431.177883138744\\
% 170	440.153827478491\\
% 180	444.494771958993\\
% 190	417.538315682876\\
% 200	366.12186312833\\
% 210	291.951401599879\\
% 220	228.073178689627\\
% 230	174.132852848014\\
% 240	135.763759762549\\
% };
%\addlegendentry{M=8, MZM\_BW=90GHz}

\addplot [color=blue, line width=1.65pt,mark=*, mark options={solid, fill=white, mark size=2.5pt}]
  table[row sep=crcr]{%
130	259.981840992615\\
140	279.972434803068\\
150	299.951162678003\\
160	319.911893913989\\
170	339.840954916746\\
180	359.66410306701\\
190	379.343238567631\\
200	398.215226080228\\
210	412.819201922187\\
220	410.552636143861\\
230	368.884576864881\\
240	310.395133996173\\
};
%\addlegendentry{M=4, MZM\_BW=100GHz}

\addplot [color=ForestGreen, line width=1.65pt, mark=*, mark options={solid, fill=white, mark size=2.5pt}]
  table[row sep=crcr]{%
130	360.091701172285\\
140	384.260577280602\\
150	411.943294544828\\
160	434.017208971923\\
170	450.270023014125\\
180	469.908470322326\\
190	473.27142623129\\
200	472.597166512396\\
210	454.495250303498\\
220	386.675439027727\\
230	306.096851315467\\
240	237.190920155785\\
};
%\addlegendentry{M=8, MZM\_BW=100GHz}

\addplot [color=blue, line width=1.65pt, dashed, mark=square*, mark options={solid, fill=white, mark size=2.5pt}]
  table[row sep=crcr]{%
130	259.980797194585\\
140	279.971678865392\\
150	299.953445243584\\
160	319.912295177895\\
170	339.845976751762\\
180	359.708469378732\\
190	379.520063617559\\
200	399.077840047496\\
210	418.038498618538\\
220	435.086727816774\\
230	443.069910929838\\
240	422.444745017061\\
};
%\addlegendentry{M=4, MZM\_BW=110GHz}

\addplot [color=ForestGreen, line width=1.65pt, dashed,mark=square*, mark options={solid, fill=white, mark size=2.5pt}]
  table[row sep=crcr]{%
130	359.45992170595\\
140	386.676264361693\\
150	410.896229996374\\
160	427.108271470289\\
170	451.726227517582\\
180	469.581365513938\\
190	484.598237412345\\
200	500.303184588899\\
210	500.501650960204\\
220	490.896433251681\\
230	446.444846354296\\
240	373.516262477755\\
};
%\addlegendentry{M=8, MZM\_BW=110GHz}

\addplot [color=blue, line width=1.65pt, dashdotted, mark=diamond*, mark options={solid, fill=white, mark size=2.5pt}]
  table[row sep=crcr]{%
130	259.97933543787\\
140	279.972644091423\\
150	299.950087237578\\
160	319.915693423313\\
170	339.840233498866\\
180	359.733817747144\\
190	379.49719190954\\
200	399.119225731351\\
210	418.56703132247\\
220	436.821988918072\\
230	453.559281135222\\
240	464.158620238878\\
};
%\addlegendentry{M=4, MZM\_BW=120GHz}

\addplot [color=ForestGreen, line width=1.65pt, dashdotted, mark=diamond*, mark options={solid, fill=white, mark size=2.5pt}]
  table[row sep=crcr]{%
130	360.678693342132\\
140	383.553205167753\\
150	407.871545995402\\
160	434.290687265769\\
170	450.547740311508\\
180	469.731877240604\\
190	484.991824619043\\
200	505.25673499181\\
210	511.214099595085\\
220	516.600273362884\\
230	512.767727820099\\
240	475.513265462729\\
};
%\addlegendentry{M=8, MZM\_BW=120GHz}

%%%%% PAM-6

% \addplot [color=red, line width=1.65pt, mark=*, mark options={solid, fill=white}]
%   table[row sep=crcr]{%
% 130	328.106010562391\\
% 140	351.69621892383\\
% 150	375.649844918512\\
% 160	396.019857922208\\
% 170	413.963622061627\\
% 180	412.423673820819\\
% 190	380.604939441097\\
% 200	297.727325999675\\
% 210	162.411828556212\\
% 220	57.0462862369013\\
% % 230	12.2509525554532\\
% % 240	0.254368705635343\\
% };
% \addlegendentry{M=6, MZM\_BW=90GHz}

\addplot [color=red, line width=1.65pt, mark=*, mark options={solid, fill=white, mark size=2.5pt}]
  table[row sep=crcr]{%
130	327.878756221023\\
140	352.026107947082\\
150	373.979467053738\\
160	397.215151527787\\
170	418.187090321318\\
180	435.087983356007\\
190	449.537032793212\\
200	440.569628539838\\
210	404.863352293156\\
220	306.805542945313\\
230	149.680503922177\\
240	58.9134710392216\\
};
% \addlegendentry{M=6, MZM\_BW=100GHz}

\addplot [color=red, line width=1.65pt, dashed,mark=square*, mark options={solid, fill=white, red, mark size=2.5pt}]
  table[row sep=crcr]{%
130	327.834913522744\\
140	352.638004697474\\
150	375.987912824489\\
160	398.720032309616\\
170	417.840397995185\\
180	437.112631758832\\
190	454.621703546542\\
200	471.714812627731\\
210	467.975063583606\\
220	448.921033334383\\
230	388.47540123157\\
240	241.245196852386\\
};
% \addlegendentry{M=6, MZM\_BW=110GHz}

\addplot [color=red, line width=1.65pt, mark=diamond*,dashdotted, mark options={solid, fill=white, mark size=2.5pt}]
  table[row sep=crcr]{%
130	328.142781352704\\
140	352.234370828796\\
150	375.561624692311\\
160	396.857849958511\\
170	419.873201839244\\
180	440.55223198532\\
190	456.802129126144\\
200	461.526928461231\\
210	476.191148157237\\
220	480.05871677488\\
230	472.886486050999\\
240	404.721472436385\\
};
% \addlegendentry{M=6, MZM\_BW=120GHz}

\draw[<->,style={draw,>=stealth,rounded corners},very thick,black] (axis cs:146,200) -- (axis cs: 200,200) node[midway,above=0pt, yshift=+.001cm, xshift=+.33cm] {\large $\mathbf{54.1}$ \textbf{GBd}};

\draw[<->,style={draw,>=stealth,rounded corners},very thick,black] (axis cs:162,260) -- (axis cs: 200,260) node[midway,above=0pt, yshift=+.001cm, xshift=-.09cm] {\large$\mathbf{38.6}$ \textbf{GBd}};

\addplot [color=mygreen,line width=1pt  ,mark=diamond*, draw=black,mark options={solid, fill=mygreen, mark size=5.5pt}]
  table[row sep=crcr]{%
145.9 400\\
};

\addplot [color=blue,line width=1pt  ,mark=diamond*, draw=black,mark options={solid, fill=blue, mark size=5.5pt}]
  table[row sep=crcr]{%
200 400\\
};

\addplot [color=red,line width=1pt  ,mark=diamond*, draw=black,mark options={solid, fill=red, mark size=5.5pt}]
  table[row sep=crcr]{%
162 400.5\\
};

\addplot [color=ForestGreen,line width=1pt  ,mark=square*, draw=black,mark options={solid, fill=ForestGreen, mark size=3.5pt}]
  table[row sep=crcr]{%
217.8 400\\
};

\addplot [color=red,line width=1pt  ,mark=square*, draw=black,mark options={solid, fill=red, mark size=3.5pt}]
  table[row sep=crcr]{%
210.5 400\\
};

\addplot [color=blue,line width=1pt  ,mark=square*, draw=black,mark options={solid, fill=blue, mark size=3.5pt}]
  table[row sep=crcr]{%
222.4 400\\
};

\draw[dashed,very thick, ForestGreen,opacity=0.7] (axis cs:145.9,0) -- (axis cs:145.9,400);

\draw[dashed,very thick, red,opacity=0.7] (axis cs:162,0) -- (axis cs:162,400);

\draw[dashed,very thick, blue] (axis cs:200,0) -- (axis cs:200,400);

% \draw[<->,style={draw,>=stealth,rounded corners},very thick,black] (axis cs:72,115) -- (axis cs: 100,115) node[midway,above=0pt, yshift=+.001cm, xshift=-.1cm] {\footnotesize $\mathbf{28}$ \textbf{GBd}};

\addplot [black, very thick, dashed, domain=0:300] {400}
    node[midway, below=-0.2pt,xshift=-2.5cm] {};

% \path[name path=axis1] (axis cs:25,400) -- (axis cs:270,400);
% \path[name path=axis2]  (axis cs:25,0) -- (axis cs:270,0);

% \addplot[gray, opacity=0.2] fill between[of=axis1 and axis2];

% \path[<->,very thick, draw, >=latex]
%     (axis cs:146,400) -- ++(10pt,0pt)
%     node[below, midway] {$ 54.1 \ \text{Gbaud}$}
%     -- (axis cs:200,400);

\end{axis}
\end{tikzpicture}%

%% file: Figures/plots/Var_code/400G/Rc_Rs_PCB_9_SNR_28_400G.tex
\definecolor{mycolor}{RGB}{255,237,160}
\definecolor{myblue}{RGB}{127,205,187}
\definecolor{mygreen}{RGB}{65,171,93}

\begin{tikzpicture}

\begin{axis}[%
width=7.5cm, 
height=6.5cm,
scale only axis,
xmin=115,
xmax=250,
xlabel style={font=\color{white!15!black}},
xlabel={\Large $R_{\text{s}}$ (Gbaud)},
% ymode=log,
ymin=0.5,
ymax=1.01,
yminorticks=true,
ylabel style={font=\color{white!15!black}},
ylabel={\Large $R_{\text{c}}$},
axis background/.style={fill=white},
title style={font=\bfseries},
% title={BER vs OMA for FFE @ 200G},
% axis x line*=bottom,
% axis y line*=left,
grid=major,
grid style={dashed,lightgray!75},
legend style={font=\large, legend cell align=left, at={(0.27,0.39)}, anchor=north east}, 
]

% \addlegendimage{color=blue, line width=1.65pt}
% \addlegendentry{PAM4}
% \addlegendimage{color=red, line width=1.65pt}
% \addlegendentry{PAM6}
% \addlegendimage{color=ForestGreen, line width=1.65pt}
% \addlegendentry{PAM8}
\addlegendimage{color=black, line width=1.65pt}
\addlegendentry{ \large $R_{\text{c}}^\text{req}$}
\addlegendimage{color=black, line width=1.65pt,densely dashdotted}
\addlegendentry{\large $R_{\text{c}}^\text{air}$}
% \addlegendimage{color=blue, line width=1.65pt}
% \addlegendentry{PAM4}
% \addlegendimage{color=red, line width=1.65pt}
% \addlegendentry{PAM6}
% \addlegendimage{color=ForestGreen, line width=1.65pt}
% \addlegendentry{PAM8}
% \addlegendimage{color=black, line width=1.65pt, mark=diamond*, mark options={solid, fill=white}}
% \addlegendentry{90 GHz}
% \addlegendimage{color=black, line width=1.65pt,dashed, mark=square*, mark options={solid, fill=white}}
% \addlegendentry{100 GHz}
% \addlegendimage{color=black, line width=1.65pt, dashdotted,mark=diamond*, mark options={solid, fill=white}}
% \addlegendentry{110 GHz}
% \addlegendimage{color=black, line width=1.65pt, dotted, mark=triangle* , mark options={solid, fill=white}}
% \addlegendentry{120 GHz}

\addplot [color=blue, line width=1.5pt]
  table[row sep=crcr]{%
130	1.53846153846154\\
140	1.42857142857143\\
150	1.33333333333333\\
160	1.25\\
170	1.17647058823529\\
180	1.11111111111111\\
190	1.05263157894737\\
200	1\\
210	0.952380952380952\\
220	0.909090909090909\\
222.4 0.898\\
};
% \addlegendentry{PAM-4}

\addplot [color=blue, line width=1.5pt,opacity=0.3]
  table[row sep=crcr]{%
222.4 0.898\\
230	0.869565217391304\\
240	0.833333333333333\\
};

% \addplot [color=blue,line width=1pt  ,mark=square*, draw=black,mark options={solid, fill=blue, mark size=3.1pt}]
%   table[row sep=crcr]{%
% 222.4 0.898\\
% };

\addplot [color=red, line width=1.5pt,opacity=0.3]
  table[row sep=crcr]{%
130	1.19031632995244\\
140	1.10529373495583\\
150	1.03160748595878\\
160	0.967132018086354\\
161 0.957\\
};

\addplot [color=red, line width=1.5pt]
  table[row sep=crcr]{%
161 0.957\\
170	0.910241899375392\\
180	0.859672904965648\\
190	0.814426962599035\\
200	0.773705614469083\\
210	0.736862489970555\\
210.5 0.734\\
};

\addplot [color=red, line width=1.5pt,opacity=0.3]
  table[row sep=crcr]{%
210.5 0.734\\
220	0.703368740426439\\
230	0.672787490842681\\
240	0.644754678724236\\
};

% \addlegendentry{PAM-6}
\addplot [color=mycolor,line width=1pt  ,mark=diamond*, draw=black,mark options={solid, fill=mycolor, mark size=5pt}]
  table[row sep=crcr]{%
161 0.957\\
};

\addplot [color=ForestGreen,opacity=0.3, line width=1.5pt]
  table[row sep=crcr]{%
130	1.02564102564103\\
140	0.952380952380952\\
145.9 0.9138\\
};
\addplot [color=ForestGreen, line width=1.5pt]
  table[row sep=crcr]{%
145.9 0.9138\\
150	0.888888888888889\\
160	0.833333333333333\\
170	0.784313725490196\\
180	0.740740740740741\\
190	0.701754385964912\\
200	0.666666666666667\\
210	0.634920634920635\\
217.8 0.612\\
};
\addplot [color=ForestGreen, line width=1.5pt,opacity=0.3]
  table[row sep=crcr]{%
217.8 0.612\\
220	0.606060606060606\\
230	0.579710144927536\\
240	0.555555555555556\\
};

% \addlegendentry{PAM-8}

% \draw[dashed,very thick,ForestGreen,opacity=1.1] (axis cs:120,0.9138) -- (axis cs:145.9,0.9138) node[midway,below=0.2pt, yshift=-.001cm, xshift=-.07cm] {};

\addplot [color=mygreen,line width=1pt  ,mark=diamond*, draw=black,mark options={solid, fill=mygreen, mark size=5pt}]
  table[row sep=crcr]{%
147.1 0.907\\
};

\addplot[
  only marks,
  mark=star*,
  mark size=7pt,
  draw=ForestGreen,
  fill=ForestGreen
] coordinates {(145.9,0.9138)};
\node[anchor=west, text=ForestGreen,yshift=-0.57cm, xshift=-1.6cm,rotate=0]
  at (axis cs:145.9,0.9138) {\large $\mathbf{R_{\text{c}}=0.91}$};

% \draw[dashdotted,very thick,red,opacity=1.1] (axis cs:120,0.957) -- (axis cs:161,0.957) node[midway,below=0.2pt, yshift=-.001cm, xshift=-.09cm] {};

\addplot [color=red,line width=1pt  ,mark=diamond*, draw=black,mark options={solid, fill=red, mark size=5pt}]
  table[row sep=crcr]{%
161 0.957\\
};

\addplot [color=blue,line width=1pt  ,mark=diamond*, draw=black,mark options={solid, fill=blue, mark size=5pt}]
  table[row sep=crcr]{%
200 1\\
};

\addplot[
  only marks,
  mark=star*,
  mark size=7pt,
  draw=ForestGreen,
  fill=ForestGreen
] coordinates {(145.9,0.9138)};
\node[anchor=west, text=blue,yshift=-.045cm, xshift=0.2cm,rotate=0]
  at (axis cs:200,0.998) {\large $\mathbf{R_{c}=1}$};

% Second point (red)
\addplot[
  only marks,
  mark=star*,
  mark size=2.5pt,
  draw=red,
  fill=red
] coordinates {(161,0.957)};
\node[anchor=west, text=red,yshift=0.1cm, xshift=+.1cm]
  at (axis cs:161,0.957) {\large $\mathbf{R_{c}=0.95}$};

% \path[name path=axis1] (axis cs:110,0) -- (axis cs:110,1);
% \path[name path=axis2]  (axis cs:134,0) -- (axis cs:134,1);

% \addplot[ForestGreen, opacity=0.2] fill between[of=axis1 and axis2];

% \path[name path=axis_1] (axis cs:134,0) -- (axis cs:134,1);
% \path[name path=axis_2]  (axis cs:156,0) -- (axis cs:156,1);

% \addplot[red, opacity=0.2] fill between[of=axis_1 and axis_2];

% \path[name path=axis_3] (axis cs:156,0) -- (axis cs:156,1);
% \path[name path=axis_4]  (axis cs:200,0) -- (axis cs:200,1);

% \addplot[blue, opacity=0.2] fill between[of=axis_3 and axis_4];

\addplot [black, thick, dashed, domain=100:260] {0.9449}
    node[midway, above=0.05pt,xshift=3.1cm] {\large  \textbf{KP4}};

% \draw[dashed,very thick,ForestGreen,opacity=0.3] (axis cs:145.9,0) -- (axis cs:145.9,0.9138);
% \draw[dashed,very thick,red,opacity=0.3] (axis cs:161,0) -- (axis cs:161,0.957);
% \draw[dashed,very thick,blue,opacity=0.3] (axis cs:200,0) -- (axis cs:200,1);
% \node[regular polygon, regular polygon sides=5,
%       draw=black, fill=black,
%       inner sep=0pt, minimum size=3pt,
%       on layer=axis foreground]
%       at (axis cs:212.5,0.9449) {} node[pos=0, below=0pt, xshift=-0.95em, yshift=1.1em,black] {$R_{C}^{Ideal}$};

%%%%%%%%%%%%%%%%%%%%%%%%%%%%%%%%%%%%%%%%%%%%%%%%%%%%%%%%%%%%%%%%%%%%%%%%%%%%%%%% Ideal code rate
%%%%%%%%%%%%%%%%%%%%%%%%%%%%%%%%%%%%%%%%%%%%%%%%%%%%%%%%%%%%%%%%%%%%%%%%%%%%%%%% %%%%%%%%%%%%%%%%%%%%%%%%%%%%%%%%%%%%%%%%%%%%%%%%%%%%%%%%%%%%%%%%%%%%%%%%%%%%%%%% 

\addplot [color=blue, densely dashdotted, line width=1.8pt,opacity=0.55]
  table[row sep=crcr]{%
130	0.999925090106609\\
140	0.999889921737166\\
150	0.999839321436006\\
160	0.999734350071302\\
170	0.999501456140664\\
180	0.999039527946726\\
190	0.998240686658091\\
200	0.995093877693586\\
210	0.980618146681024\\
220	0.932587283504964\\
230	0.796894177733778\\
240	0.652836032072853\\
};

\addplot [color=blue,line width=1pt  ,mark=square*, draw=black,mark options={solid, fill=blue, mark size=3.1pt}]
  table[row sep=crcr]{%
222.4 0.898\\
};

\addplot[
  only marks,
  mark=star*,
  mark size=2.5pt,
  draw=red,
  fill=red
] coordinates {(161,0.957)};
% \node[anchor=west, text=blue,yshift=-0.3cm, xshift=-1.2cm]
%   at (axis cs:222.4,0.898) {\normalsize $\mathbf{r \approx 0.89}$};

\addplot [color=red, densely dashdotted, line width=1.8pt,opacity=0.55]
  table[row sep=crcr]{%
130	0.975698594435943\\
140	0.972730629136989\\
150	0.964500044518774\\
160	0.960398727778865\\
170	0.951628528470864\\
180	0.935083376418261\\
190	0.915287700483896\\
200	0.852177987914578\\
210	0.745821544671402\\
220	0.539493570743236\\
% 230	0.251757926654674\\
% 240	0.0949618402310572\\
};

\addplot [color=red,line width=1pt  ,mark=square*, draw=black,mark options={solid, fill=red, mark size=3.1pt}]
  table[row sep=crcr]{%
210.5 0.734\\
};

\addplot[
  only marks,
  mark=star*,
  mark size=2.5pt,
  draw=red,
  fill=red
] coordinates {(161,0.957)};
% \node[anchor=west, text=red,yshift=-0.16cm, xshift=.02cm]
%   at (axis cs:210,0.73) {\normalsize $\mathbf{r \approx 0.73}$};

\addplot [color=ForestGreen, densely dashdotted, line width=1.8pt, opacity=0.55]
  table[row sep=crcr]{%
130	0.922350178925312\\
140	0.919494273241599\\
150	0.902499014895591\\
160	0.894699485928916\\
170	0.886614390457498\\
180	0.860378748380459\\
190	0.847865165765655\\
200	0.780166105771785\\
210	0.71233002881184\\
220	0.584691585215726\\
230	0.447648143508548\\
% 240	0.331480287548269\\
};

\addplot [color=ForestGreen,line width=1pt  ,mark=square*, draw=black,mark options={solid, fill=ForestGreen, mark size=3.1pt}]
  table[row sep=crcr]{%
217.8 0.612\\
};

\addplot[
  only marks,
  mark=star*,
  mark size=2.5pt,
  draw=red,
  fill=red
] coordinates {(161,0.957)};
% \node[anchor=west, text=ForestGreen,yshift=-0.25cm, xshift=.1cm]
  % at (axis cs:217.8,0.612) {\normalsize $\mathbf{r \approx 0.61}$};

\draw[dashed,very thick, ForestGreen,opacity=0.7] (axis cs:147.1,0) -- (axis cs:147.1,0.91);

\draw[dashed,very thick, red,opacity=0.7] (axis cs:161,0) -- (axis cs:161,0.95);

\draw[dashed,very thick, blue,opacity=0.7] (axis cs:200,0) -- (axis cs:200,1);
% ] coordinates {(212.5,0.9449)};
% \node[anchor=west, text=red,yshift=8pt, xshift=+.01cm]

\draw[<->,style={draw,>=stealth,rounded corners},very thick,black] (axis cs:161,0.61) -- (axis cs: 200,0.61) node[midway,above=0pt, yshift=+.001cm, xshift=-.09cm] {\large $\mathbf{38.6}$ \textbf{GBd}};

\draw[<->,style={draw,>=stealth,rounded corners},very thick,black] (axis cs:147.1,0.55) -- (axis cs: 200,0.55) node[midway,above=0pt, yshift=+.001cm, xshift=+.26cm] {\large $\mathbf{54.1}$ \textbf{GBd}};

% \draw[<->,style={draw,>=stealth,rounded corners},very thick,ForestGreen] (axis cs:190,0.70175438596491) -- (axis cs: 190,0.847865165765655) node[midway,above=0pt, yshift=+1.001cm, xshift=-.16cm,rotate=90] {\large \textbf{Max Back-off}};

\draw[<->,style={draw,>=stealth,rounded corners},
very thick,ForestGreen]
(axis cs:190,0.70175438596491) --
(axis cs:190,0.847865165765655)
node[midway,
above=0pt,
yshift=+0.151cm,
xshift=-.21cm,
rotate=90,
align=center]
{\large \textbf{Max}};

\draw[<->,style={draw,>=stealth,rounded corners},
very thick,line width=1.9pt,ForestGreen]
(axis cs:190,0.70175438596491) --
(axis cs:190,0.847865165765655)
node[midway,
above=0pt,
yshift=+0.401cm,
xshift=0.05cm,
rotate=90,
align=center]
{\large \textbf{rate backoff}};

% 190	0.847865165765655\\
% 190	0.701754385964912\\

%     \raisebox{0pt}{\tikz\node[draw,fill=white,draw=black,rotate=45,inner sep=0pt,minimum size=5pt]{};} &
%     {\color{blue} 1.00} & {\color{red} 0.95} & {\color{ForestGreen} 0.91}\\
%     % row: square
%     \raisebox{0pt}{\tikz\node[draw,fill=white,draw=black,inner sep=0pt,minimum size=5pt]{};} &
%     {\color{blue} 0.89} & {\color{red} 0.73} & {\color{ForestGreen} 0.61}\\
%   \end{tabular}%
% };

\end{axis}
\end{tikzpicture}%

%% file: ICTON_final_standalone/Main_Text.bbl
\begin{thebibliography}{1}
\providecommand{\url}[1]{#1}
\csname url@samestyle\endcsname
\providecommand{\newblock}{\relax}
\providecommand{\bibinfo}[2]{#2}
\providecommand{\BIBentrySTDinterwordspacing}{\spaceskip=0pt\relax}
\providecommand{\BIBentryALTinterwordstretchfactor}{4}
\providecommand{\BIBentryALTinterwordspacing}{\spaceskip=\fontdimen2\font plus
\BIBentryALTinterwordstretchfactor\fontdimen3\font minus \fontdimen4\font\relax}
\providecommand{\BIBforeignlanguage}[2]{{%
\expandafter\ifx\csname l@#1\endcsname\relax
\typeout{** WARNING: IEEEtran.bst: No hyphenation pattern has been}%
\typeout{** loaded for the language `#1'. Using the pattern for}%
\typeout{** the default language instead.}%
\else
\language=\csname l@#1\endcsname
\fi
#2}}
\providecommand{\BIBdecl}{\relax}
\BIBdecl

\bibitem{DI_CHE_IMDD_200}
D.~Che and X.~Chen, ``Modulation {{Format}} and {{Digital Signal Processing}} for {{IM-DD Optics}} at {{Post-200G Era}},'' \emph{JLT}, vol.~42, no.~2, pp. 588--605, Jan. 2024.

\bibitem{st_arnaultNet32Tbps_OFC2025}
C.~St-Arnault, S.~Bernal, D.~Kita, R.~Dickson, M.~Y. Abdelaziz, A.~Nikic, B.~Qiu, B.~Krueger, F.~Pittal\`{a}, C.~Reimer, B.~Beggs, N.~Ben-Hamida, and D.~V. Plant, ``{Net 3.2 Tbps 225 Gbaud PAM4 O-Band IM/DD 2 km Transmission Using FR8 and DR8 with a CMOS 3 nm SerDes and TFLN Modulators},'' in \emph{OFC}.\hskip 1em plus 0.5em minus 0.4em\relax Optica Publishing Group, 2025, p. Th4B.1.

\bibitem{Pang_200G}
X.~Pang, O.~Ozolins, R.~Lin, L.~Zhang, A.~Udalcovs, L.~Xue, R.~Schatz, U.~Westergren, S.~Xiao, W.~Hu, G.~Jacobsen, S.~Popov, and J.~Chen, ``{200 Gbps/Lane IM/DD Technologies for Short Reach Optical Interconnects},'' \emph{JLT}, vol.~38, no.~2, pp. 492--503, 2020.

\bibitem{agrawalFiberopticCommunicationSystems2002}
G.~P. Agrawal, \emph{{Fiber-Optic Communication Systems}}.\hskip 1em plus 0.5em minus 0.4em\relax John Wiley \& Sons, 2012.

\bibitem{farhood_P802_3dj}
A.~Farhood \emph{et~al.}, ``Concatenated {FEC} baseline proposal for 200~{Gb/s} per lane {IM-DD} optical {PMD},'' {IEEE P802.3dj Ethernet Task Force}, Feb. 2023.

\bibitem{PAM_6}
A.~Uchiyama, S.~Okuda, T.~Tsuji, Y.~Hokama, M.~Shirao, K.~Abe, T.~Yamatoya, and Y.~Yamauchi, ``{Demonstration of 155 Gbaud PAM4 and PAM6 EML with Narrow High-Mesa EA Modulator for 400 Gbps per Lane Transmission},'' in \emph{OFC}, 2024, pp. 1--3.

\bibitem{PAM_8}
M.~S.-B. Hossain, J.~Wei, F.~Pittalà, N.~Stojanović, S.~Calabrò, T.~Rahman, T.~Wettlin, C.~Xie, M.~Kuschnerov, and S.~Pachnicke, ``{Single-Lane 402 Gb/s PAM-8 IM/DD Transmission Based on a Single DAC and an O-Band Commercial EML},'' in \emph{OECC}, 2021, pp. 1--3.

\bibitem{luo800GLinearDirect}
F.~Luo, L.~Fang, C.~Cheng, L.~Guo, D.~Shen, H.~Kang, A.~Sun, Z.~Su, and X.~Jiang, ``{800G Linear Direct Drive Network System Design \& Implementation},'' in \emph{Proceedings of DesignCon 2024}.\hskip 1em plus 0.5em minus 0.4em\relax Santa Clara, CA, USA: DesignCon, 2024.

\bibitem{achievable_rate_HD}
A.~Sheikh, A.~G.~i. Amat, and G.~Liva, ``Achievable information rates for coded modulation with hard decision decoding for coherent fiber-optic systems,'' \emph{JLT}, vol.~35, no.~23, pp. 5069--5078, 2017.

\end{thebibliography}
